\documentclass[preprint2]{aastex62}

\def\Mgii{Mg\,{\sc ii}}

\def\Cii{C\,{\sc ii}}
\def\Ci{C\,{\sc i}}

\shorttitle{The Host Galaxy of The Most Massive SMBH at $z=6.327$}
\shortauthors{Wang et al.}

\begin{document}

\title{Spatially Resolved Interstellar Medium and Highly-Excited Dense Molecular Gas in the Most Luminous Quasar at $z=6.327$}

\correspondingauthor{Feige Wang}
\email{fgwang@physics.ucsb.edu}

\author[0000-0002-7633-431X]{Feige Wang}
\affil{Department of Physics, University of California, Santa Barbara, CA 93106-9530, USA}
\affil{Kavli Institute for Astronomy and Astrophysics, Peking University, Beijing 100871, China}

\author{Ran Wang}
\affil{Kavli Institute for Astronomy and Astrophysics, Peking University, Beijing 100871, China}

\author[0000-0003-3310-0131]{Xiaohui Fan}
\affil{Steward Observatory, University of Arizona, 933 North Cherry Avenue, Tucson, AZ 85721, USA}

\author[0000-0002-7350-6913]{Xue-Bing Wu}
\affil{Kavli Institute for Astronomy and Astrophysics, Peking University, Beijing 100871, China}
\affil{Department of Astronomy, School of Physics, Peking University, Beijing 100871, China}

\author[0000-0001-5287-4242]{Jinyi Yang}
\affil{Steward Observatory, University of Arizona, 933 North Cherry Avenue, Tucson, AZ 85721, USA}

\author{Roberto Neri}
\affil{Institute de Radioastronomie Millimetrique, St. Martin d'Heres, F-38406, France}

\author[0000-0002-5367-8021]{Minghao Yue}
\affil{Steward Observatory, University of Arizona, 933 North Cherry Avenue, Tucson, AZ 85721, USA}

\begin{abstract}

Among more than 200 quasars known at $z\gtrsim6$, only one object, J0100+2802 (z=6.327), was found hosting a $>10^{10}M_\odot$ super-massive black hole (SMBH). In order to investigate the host galaxy properties of J0100+2802, we performed multi-band ALMA observations, aiming at mapping the dust continuum,  [\Cii] and CO(6-5) emission lines with sub-kiloparsec scale resolution, as well as detecting high-J CO lines in CO(11-10), CO(10-9), and CO(7-6). The galaxy size is measured to be $R_{\rm major}=3.6\pm0.2$ kpc from the high resolution continuum observations. No ordered motion on kilo-parsec scales was found in both [\Cii] and CO(6-5) emissions. The velocity dispersion is measured to be 161$\pm$7 km s$^{-1}$, $\sim$3 times smaller than that estimated from the local M-$\sigma$ relation. In addition, we found that the CO emission is more concentrate (a factor of 1.8$\pm$0.4) than the [\Cii] emission. Together with CO(2-1) detected by VLA, we measured the CO Spectral Line Energy Distribution (SLED), which is best fit by a two-components model, including a cool component at $\sim24$ K with a density of $n_{\rm (H_2)}$=10$^{4.5}$ cm$^{-3}$, and a warm component at $\sim224$ K with a density of $n_{\rm (H_2)}$=10$^{3.6}$ cm$^{-3}$. We also fitted the dust continuum with a graybody model, which indicates that it has either a high dust emissivity $\beta\gtrsim2$ or a hot dust temperature $T_{\rm dust}\gtrsim60$ K, or a combination of both factors. The highly excited CO emission and hot dust temperature suggest that the powerful AGN in J0100+2802 could contribute to the gas and dust heating although future observations are needed to confirm this. 
\end{abstract}


\section{Introduction} \label{sec_intro}
In the past decade, the advent of wide area optical and infrared surveys has resulted in the discovery of more than 200 luminous quasars at $z\sim6$ \citep[e.g.][]{Fan01,Jiang08,Willott10,Mortlock11,Banados16,Wang17,Wang19,Matsuoka18,Yang18} with the highest redshift at $z=7.5$ \citep{Banados18}. These quasars are powered by $\sim10^{9}$ $M_\odot$ supermassive black holes \citep[SMBHs; e.g.][]{Jiang07,Shen18}.  The discovery of a $1.24\times10^{10}$ $M_\odot$ BH at $z=6.3$ \citep{Wu15} and $\sim10^{9}$ $M_\odot$ BHs at $z>7$ \citep{Mortlock11,Banados18,Wang18,Yang18} challenges theories of the formation and growth of SMBHs in the early universe \citep[e.g.][]{Pacucci15}. 

Deep imaging of quasar host galaxies allows studies of how mergers and star formation activities affect the central nuclei, and vice versa. However, the nuclear region of quasar is extremely bright in the rest-frame ultraviolet (UV), which prevents direct detections of the host galaxies and galactic environments of luminous quasars with ground based telescope. \cite{Mechtley12} obtained deep near-infrared {\it Hubble Space Telescope} Wide Field Camera 3 images of the $z=6.42$ quasar J1148+5251. However, even with careful point-spread function (PSF) subtraction, only an upper limit of UV emissions from the host galaxy is measured.  Such observations indicate an infrared excess of log(IRX)$>$1.0 in the host galaxies of luminous quasars at the end of reionization, comparable to that of the most luminous infrared galaxies in the local universe. 

On the other hand, sub-millimeter and millimeter observations of both the dust continuum, [\Cii] 158$\mu$m fine structure lines and molecular CO lines of high redshift quasar host galaxies directly probe the star formation and interstellar medium properties of the quasar hosts. This provides the most useful observational probe of the growth of massive galaxies and the relation between SMBHs and their hosts in early epochs \citep[See][and references therein]{Carilli13,Gallerani17}. The recent [\Cii] survey of $\sim$30 $z\gtrsim$6 quasars with the Atacama Large Millimeter Array (ALMA) shows that the SMBH-host galaxy mass ratio on average is about one dex above the local value \citep{Decarli18}, and that there is no correlation between the BH accretion measured from UV luminosity and stellar mass and star formation inferred from far-infrared (FIR) luminosity \citep{Venemans18}. These findings suggest that BHs in the most luminous quasars at the end of reionization might do not co-evolve with their host galaxies, in the sense that they do not following the same relation between the mass of SMBHs and the bulge masses of their host galaxies established in the local universe.

The ultra-luminous quasar, SDSS J010013.02 +280225.8 (hereafter, J0100+2802) at $z=6.3$, with a bolometric luminosity of $4.29\times 10^{14}$ $\rm L_\odot$ and a $1.24\times10^{10}$ $M_\odot$ SMBH, is by far the most optically luminous quasar with the most massive SMBH known at $z>6$.
J0100+2802 was discovered by \cite{Wu15} using optical plus mid-infrared color selection developed by \cite{Wang16}. 
The sphere of influence ($r_{\rm SOI}$ = GM$_{\rm BH}$/$\sigma^2$) of such massive SMBH is at sub kilo-parsec (kpc) scale, which could be resolved by ALMA. Thus, J0100+2802 is a unique source for the study of relation between BH growth and galaxy assembly in the early universe.   

In \cite{Wangr16}, we detected [\Cii], CO(6-5) and CO(2-1) emissions in the host galaxy of J0100+2802. Our observations indicated that J0100+2802 only has  moderate FIR emission and the narrow emission line width suggests that it is likely to be well above the local M-$\sigma$ relation. However, large uncertainties remain in determining the properties of the host galaxy of J0100+2802,  limited by the low spatial resolution ($\sim$2\farcs0 or $\sim$10 kpc at the quasar redshift) and small number of CO lines detected. In this paper, we present  ALMA imaging in [\Cii] and CO(6-5) lines with sub-kpc spatial resolution. 
We also report the detection of multiple high--$J$ CO lines with ALMA observations. The [\Cii] observations were performed in ALMA Cycle 3 and the CO line observations were obtained in Cycle 5. 
The paper is organized as following: 
In \S \ref{sec_obs} we describe our new ALMA observations of J0100+2802 and basic measurements based on these observations. 
In \S \ref{sec_dust} we describe its dust properties (including dust temperature, emissivity index and dust mass), FIR luminosity and star formation rate (SFR).
In \S \ref{sec_excitation} we present constraints on its gas excitations from the spectral line energy distribution (SLED) of the CO emission lines and fine structure line ratios.
In \S \ref{sec_kinematics} we will discuss the gas kinematics, dynamical mass measurements, and mass budget in J0100+2802. We also briefly discuss the sphere of influence of the central SMBH.
We summarize our findings in \S \ref{sec_summary}. 
Throughout the paper, we adopt a $\Lambda$CDM cosmological model with parameters $\Omega_{\Lambda}$ = 0.7, $\Omega_{m}$ = 0.3, and H$_{0}$ = 70 $\rm km s^{-1} Mpc^{-1}$.

\begin{figure*}
\centering
\includegraphics[width=1.0\textwidth]{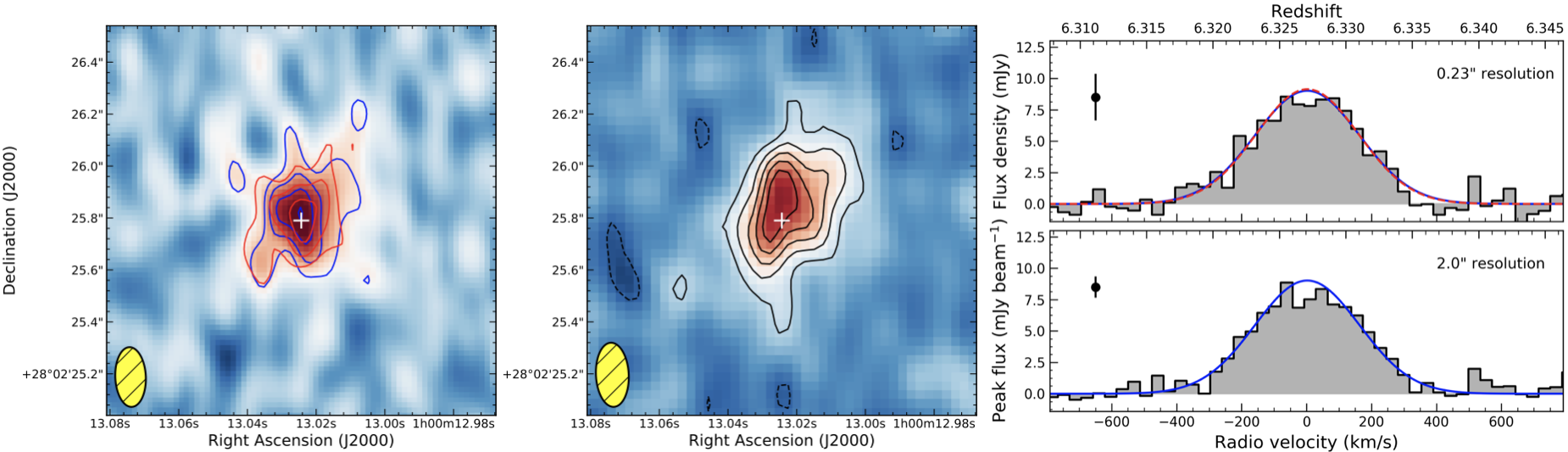}
\caption{{\it Left:} Continuum subtracted map of [\Cii] emission. The blue and red sides of the emission line are shown in blue and red contours, respectively. The blue side emission is averaged over from --300 km s$^{-1}$ to --150 km s$^{-1}$, and the red sides emission is averaged over from +150 km s$^{-1}$ to +300 km s$^{-1}$. Contour levels are +3$\sigma$, +4$\sigma$, +5$\sigma$, and +6$\sigma$.
The ALMA beam is shown in the bottom left corner, which is 0\farcs23$\times$0\farcs12. 
The white signs in all maps in this paper indicate the optical position of the quasar obtained from GAIA. 
{\it Middle}: Map of 251 GHz continuum emission. Contour levels are --2$\sigma$, +3$\sigma$, +5$\sigma$, +7$\sigma$, +9$\sigma$, and +11$\sigma$, with $\sigma\sim17\mu$Jy.  The beam size is 0\farcs25$\times$0\farcs13.
{\it Bottom Right:} Spectrum extracted from the peak pixel in the tapered [\Cii] cube. The solid blue line denotes the Gaussian fit. 
{\it Top Right:} Spectrum extracted from the full resolution [\Cii] map with an extraction aperture radius of 0\farcs67. The red dashed line is the Gaussian fit to this spectrum and the blue solid line is the Gaussian fit to the spectrum shown in the top panel. 
The bins are 31.25 MHz wide, which corresponds to $\sim$36 km s$^{-1}$. The typical 1$\sigma$ uncertainties per 31.25 MHz bin are 0.85 mJy beam$^{-1}$ and 0.21 mJy beam$^{-1}$ for the tapered [\Cii] cube and the full resolution [\Cii] cube, respectively. The uncertainty shown on top panel is the 1$\sigma$ flux density uncertainty within extraction region per 31.25 MHz, which is 1.86 mJy.
\label{fig_cii_blue_red}}
\end{figure*}

\section{Observations and Measurements} \label{sec_obs}
\subsection{[\Cii] Fine Structure Line Observations}\label{subsec_cii}
At $z=6.32$, the [\Cii] emission line is redshifted from 1901 GHZ to 259 GHz. We observed the [\Cii] emission of J0100+2802 in ALMA band-6 with the C36-6 configuration on 2016 September 4. The total on-source time is 74 minutes. 
We tuned the receivers to cover the redshifted [\Cii] line in spectra window 1 and to cover the continuum with the other three spectral windows. 
Bandpass calibration was performed through observations of J0237+2848, and J2253+1608. For the flux and amplitude calibration, the sources J0238+1636, and J2253+1608 were observed. The source J0057+3021 was observed as phase calibrator. 

The data is reduced using the Common Astronomy Software Application \citep[CASA;][]{McMullin07}, following standard reduction steps. We first subtract the underlying continuum of [\Cii] by fitting a UV-plane model with the {\it uvcontsub} task. We then image the line data cube using the {\it tclean} task with briggs weighting and a robustness parameter of 0.5, which optimizes the noise per frequency bin and the resolution of the resulting map. The synthesized beam size at the frequency of [\Cii] is 0\farcs23$\times$0\farcs12. The 1$\sigma$ root mean square (rms) sensitivity is 0.21 mJy/beam per $\sim$36 km/s channel.

We use the data cube of [\Cii] over the velocity range from --300 km s$^{-1}$ to 300 km s$^{-1}$ to derive the intensity map of [\Cii] emission. Similarly, the intensity map for all other lines described in the following sections are also derived with the same parameters. The [\Cii] emission line in J0100+2802 is clearly detected in the continuum-subtracted [\Cii] map (Figure \ref{fig_cii_blue_red}). The underlying continuum map is also shown in Figure \ref{fig_cii_blue_red}. We fit the velocity integrated [\Cii] emission intensity map with {\it imfit} and derive the image component size to be 0\farcs47$\times$0\farcs39, which is more than two times larger than the beam size in both major and minor axis direction. We measure the de-convolved image size of (0\farcs43 $\pm$ 0\farcs10)$\times$ (0\farcs34 $\pm$ 0\farcs09). Throughout this paper unless otherwise indicated, the sizes are quoted as FWHM size and the errors are estimated from the {\it imfit} fitting of elliptical Gaussians based on the work of \cite{Condon97}.
The line flux measured from 2D Gaussian fitting on the intensity map is 3.04$\pm$0.48 Jy km s$^{-1}$. Our deep ALMA imaging with sub-kpc scale (1.27$\times$0.67 kpc$^2$) resolution has well resolved the [\Cii] emission of J0100+2802. 
One way to address whether the size measured from the image plane is impacted by the `missing' flux is to fit the {\it uv}-data directly \citep[e.g.][]{Ikarashi15,Hodge16}. 
We fit the {\it uv}-data with the {\it uvmodelfit} routine in CASA by assuming an elliptical Gaussian profile. In the fitting, we only select visibility from the continuum subtracted data with frequency between 250 GHz and 250.8 GHz to ensure that we measure the size of the [\Cii] emission. The best-fit {\it uv}-model of the [\Cii] emission gives a size of (0\farcs49 $\pm$ 0\farcs06)$\times$ (0\farcs39 $\pm$ 0\farcs05) or (2.72$\pm$0.33)$\times$(2.16$\pm$0.28) kpc$^2$, which agrees with (within the uncertainties) that derived from the fitting in the image plane, suggesting that the sizes measured in the image plane are robust, and that they are not significantly affected by the presence of potentially `missing' emission.

To account for spatially extended low level [\Cii] emission and measure the total flux for [\Cii], we tapered the [\Cii] map by setting {\it uvtaper}=1\farcs5 in the {\it tclean} task. The tapered cube has a beam size of 2\farcs01$\times$1\farcs53, comparable with our previous PdBI observation \citep{Wangr16}. Figure \ref{fig_cii_blue_red} shows the spectra extracted from both the tapered  and the full resolution [\Cii] map as well as the Gaussian fits. The spectrum extracted from the tapered map uses the peak pixel in the cube and the spectrum extracted from the full resolution cube uses the flux densities within an aperture radius of 0\farcs67 centered at the center position of velocity integrated [\Cii] map. 
The extraction aperture is determined by matching the [\Cii] line flux to that of the tapered image.

The redshifts measured from both spectra are 6.3270$\pm$0.0005. We refer $z=6.327$ as the systemic redshift of J0100+2802 throughout the paper. 
The FWHM of [\Cii] line measured to be 380$\pm$16 km s$^{-1}$ from the tapered spectrum, and  373$\pm$25 km s$^{-1}$ from the full resolution spectrum. The total [\Cii] line flux is measured to be 3.64$\pm$0.22 Jy km s$^{-1}$ with a single Gaussian fitting to the spectrum extracted from the tapered cube. This is consistent with the flux (3.36$\pm$0.46 Jy km s$^{-1}$) measured from our previous PdBI observations \citep{Wangr16}. 
The line flux measured from the 2D Gaussian fitting on the intensity map is lower than that measured from the tapered spectrum, which also supports that extended emission exists beyond the bright core. 

In Figure \ref{fig_cii_blue_red} we also over-plot the contours of [\Cii] intensity from the blue and red sides of the emission line, which are derived by integrating blue side emissions from --300 km s$^{-1}$ to --150 km s$^{-1}$ and integrating red side emissions from +150 km s$^{-1}$ to +300 km s$^{-1}$, respectively. The peak position of both blue and red side emissions coincide with that of the entire [\Cii] intensity map and quasar optical position, which suggests that [\Cii] emission traced gas does not show ordered motions on scales of $\sim$1 kpc.

\begin{figure*}
\centering
\includegraphics[width=1.0\textwidth]{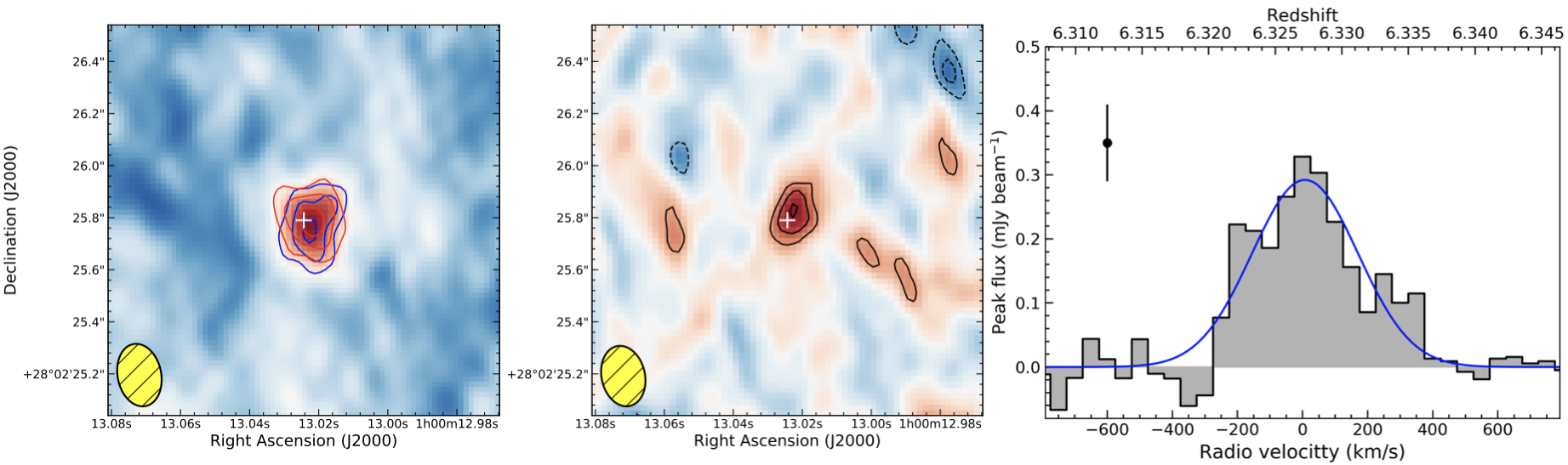}
\caption{{\it Left:} Continuum subtracted high spatial resolution map of CO(6-5) emission. The blue and red sides of the emission line are shown in cyan and red contours, respectively. The blue sides emission is averaged over from --300 km s$^{-1}$ to --150 km s$^{-1}$, and the red sides emission is averaged over from +150 km s$^{-1}$ to +300 km s$^{-1}$. Contour levels are +3$\sigma$, and +4$\sigma$.
The beam is shown in the bottom left corner, which is 0\farcs24$\times$0\farcs17. 
{\it Middle:} High resolution map of 99.4 GHz continuum emission. Contour levels are --3$\sigma$, --2$\sigma$, +2$\sigma$, +3$\sigma$, and +4$\sigma$, with $\sigma\sim6\mu$Jy.
The beam with a size of 0\farcs24$\times$0\farcs17 is plotted in the bottom left corner.
{\it Right:} Spectrum extracted from peak pixel in the CO(6-5) data cube. The solid blue line denote the Gaussian fit. The bins are 60 km s$^{-1}$ wide. The uncertainty shown on the plot is  1$\sigma$ in flux density per 60 km s$^{-1}$ bin width, which is 0.06 mJy beam$^{-1}$.
\label{fig_co65}}
\end{figure*}
 
\subsection{CO Molecular Lines Observations}
We observed CO(6-5) in ALMA band-3 with the C43-7 configuration to reach similar spatial resolution with that of the  [\Cii] observations. The data were taken from 2017 November 18 to November 20 with total on-source time of 141 minutes. 
Bandpass calibration was performed through observations of J0238+1636, and J2253+1608. For the flux and amplitude calibration, the sources J0238+1636, and J2253+1608 were observed. Same with the [\Cii] observations, the source J0057+3021 was served as phase calibrator. 
The data was processed using similar strategy for [\Cii] observations. The main difference is that we image the CO line data cubes using a weighting factor of robust$=$2 (i.e., natural weight scheme) to maximize the signal-to-noise ratio. The synthesized beam size at the frequency of CO(6-5) is 0\farcs24$\times$0\farcs17.

We fit the CO(6-5) intensity map (Figure \ref{fig_co65}) with {\it imfit}. The fitting gives the image size (convolved with beam) of 0\farcs34$\times$0\farcs27,  which is slightly larger than the beam size (0\farcs24$\times$0\farcs17). This suggests that we marginally resolved the CO(6-5) emission at the designed resolution. The de-convolved size of CO(6-5) is (0\farcs26$\pm$0.09)$\times$(0\farcs19$\pm$0.11), or (1.44$\pm$0.50)$\times$(1.05$\pm$0.61) kpc$^2$. 
As indicated in \S \ref{subsec_cii}, the size measured from image plane is robust, and also the CO(6-5) emission is only marginally resolved and thus less affected by possible `missing' fluxes, we adapt the size measured from {\it imfit} as the size of CO(6-5) emission.
The size (major axes) of CO(6-5) emission is $1.8\pm0.4$ times smaller than that of [\Cii] emission, indicating that the molecular gas is more centrally concentrated than [\Cii]. Similar result is also reported in the host galaxy of a bright $z=6$ quasar \citep{Feruglio18}.

\begin{figure*}
\centering
\includegraphics[width=1.0\textwidth]{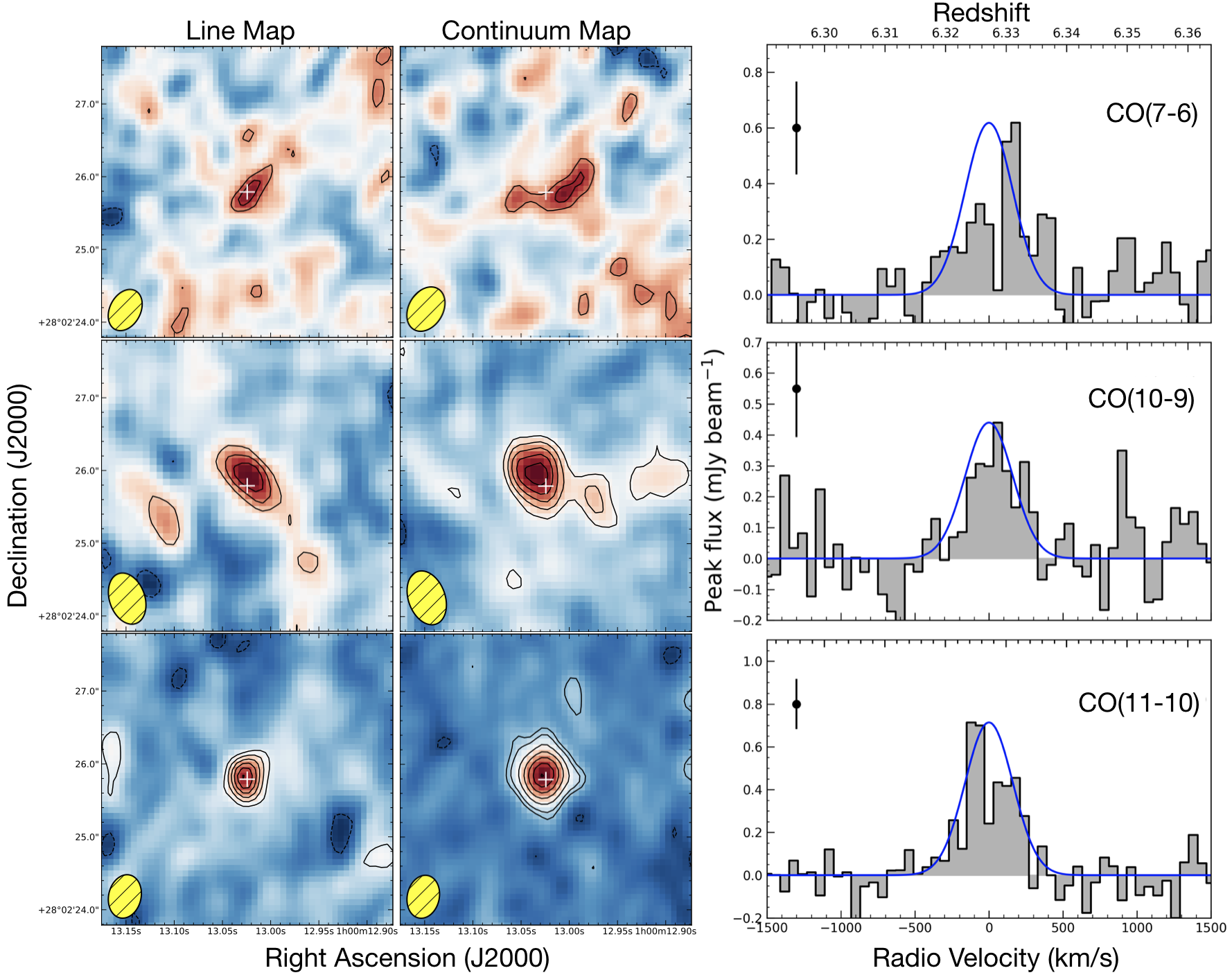} 
\caption{Low spatial resolution CO and underlying continua observations. 
The left column shows the CO(7-6), CO(10-9), and CO(11-10) continuum subtracted line emission maps from top to bottom, respectively. The contours are started from +2$\sigma$ increasing by $\sigma$, with $\sigma\sim0.06$ Jy km s$^{-1}$ beam$^{-1}$, $\sim0.05$ Jy km s$^{-1}$ beam$^{-1}$, and $\sim0.04$ Jy km s$^{-1}$ beam$^{-1}$ for CO(7-6), CO(10-9), and CO(11-10) line maps, respectively. The --2$\sigma$ contours are shown as dashed lines. 
The beams are shown as yellow ellipticals. 
The middle column shows the 103 GHz, 152 GHz, and 180 GHz continuum emissions from top to bottom, respectively. The contour levelss for the 103 GHz continuum are --2$\sigma$, +2$\sigma$, and +3$\sigma$, with $\sigma\sim12\mu$Jy beam$^{-1}$.  The contour levels for the 152 GHz continuum are --2$\sigma$, +2$\sigma$, +3$\sigma$, +4$\sigma$, +5$\sigma$, and +6$\sigma$, with $\sigma\sim17\mu$Jy beam$^{-1}$. The contour levels for the 180 GHz continuum are --2$\sigma$, +2$\sigma$, +3$\sigma$, +5$\sigma$, +7$\sigma$, +9$\sigma$, +11$\sigma$, and +13$\sigma$, with $\sigma\sim 19\mu$Jy beam$^{-1}$. 
The right column shows the spectra for CO(7-6), CO(10-9), and CO(11-10) lines from top to bottom, respectively. The blue curves are Gaussian profiles with amplitudes equal to the maximum peak flux densities of each line and FWHM=380 km s$^{-1}$ centered at the [\Cii] based redshift.  The line fluxes derived from Gaussian profiles are 0.25 Jy km s$^{-1}$, 0.18 Jy km s$^{-1}$ and 0.29 Jy km s$^{-1}$ for CO(7-6), CO(10-9), and CO(11-10), respectively.
\label{fig_coall}}
\end{figure*}

We also extract the spectrum at the peak pixels in the CO(6-5) data cube, shown in Figure \ref{fig_co65}. We fit the extracted spectrum with a single Gaussian function, which yields a redshift of 6.3271$\pm$0.0006, FWHM of 383$\pm$32 km s$^{-1}$ and line flux of 0.11$\pm$0.02 Jy km s$^{-1}$. Both redshift and line widths of CO(6-5) lines are well consistent with that of [\Cii] emission line. The line flux estimated from the extracted spectrum at peak position is about three times lower than that from the 2D Gaussian fitting of CO(6-5) intensity map. This agrees with the argument of that we marginally resolved the CO(6-5) emission;  such disparity between line fluxes measured from fitting extracted spectrum and 2D map is often seen in the marginally resolved observations of [\Cii] emission lines \citep[e.g.][see their Figure 6]{Decarli18}. In order to measure the total line flux, we tapered the CO(6-5) data cube with {\it uvtaper}=1\farcs5. We found a line flux of 0.26$\pm$0.05 Jy km s$^{-1}$ at the peak position, which is consistent with that measured in our low resolution PdBI observations \citep{Wangr16}. 

Following the analysis performed on [\Cii] emission, we derive the blue and red intensity maps by integrating the blue side emissions from --300 km s$^{-1}$ to --150 km s$^{-1}$ and the red side emissions from +150 km s$^{-1}$ to +300 km s$^{-1}$. There is no measurable offset between the blue and red side emissions of CO(6-5) line considering the beam size and the low detection significance of the blue and red side emissions (Figure \ref{fig_co65}), which suggests that CO(6-5) also does not show ordered motions on scales of $\sim$1 kpc.

We also observed CO(7-6) in ALMA band-3 with the C43-5 configuration, CO(10-9) in ALMA band-4 with the C43-4 configuration, and CO(11-10) in ALMA band-5 with the C43-5 configuration. The data were taken on 2017 December 25 with 49 minutes on-source exposure for CO(7-6), on 2018 January 15 with 49 minutes on-source exposure for CO(10-9), and on 2018 September 18 with 61 minutes on-source exposure for CO(11-10). 
Bandpass calibration was performed through observations of J0237+2848, J0238+1636, and J2253+1608. For the flux and amplitude calibration, the sources J0237+2848, J0238+1636, and J2253+1608 were observed. The source J0057+3021 was served as phase calibrator for all CO observations.
We tuned one spectral window centered at the expected frequency of each line and the other three spectral windows for the continuum. The synthesized beam size at the frequency of CO(7-6), CO(10-9) and CO(11-10) are 0\farcs60$\times$0\farcs42, 0\farcs73$\times$0\farcs49, and 0\farcs60$\times$0\farcs45, respectively.

The velocity integrated intensity map of CO(11-10), CO(10-9), and CO(7-6) are shown in Figure \ref{fig_coall}. The CO(11-10) is detected at 7$\sigma$ level with the line flux of 0.28$\pm$0.04 Jy km s$^{-1}$ at the peak position. We also use {\it imfit} to fit the line emission which gives a line flux of 0.30$\pm$0.04 Jy km s$^{-1}$, consistent with the peak line flux. The CO(10-9) and CO(7-6) emissions are marginally detected in the velocity integrated intensity map. We thus adapt the line fluxes at the peak positions of these two CO lines as their brightness, which are 0.25$\pm$0.05 Jy km s$^{-1}$ and 0.22$\pm$0.06 Jy km s$^{-1}$ for CO(10-9) and CO(7-6), respectively. To be consistent with other CO line flux measurements, we also use the line flux (0.28$\pm$0.04 Jy km s$^{-1}$) at the peak position for CO(11-10) in the following sections. We also search for  the [\Ci](2--1) 370 $\mu$m emission line (close to CO(7-6) emission) in the CO(7-6) data cube. We integrate the data over the velocity range from --300 km s$^{-1}$ to 300 km s$^{-1}$. It yields a 3-$\sigma$ upper limit line flux of 0.18 Jy km s$^{-1}$. The line maps, underlying continuum maps and the extracted spectra for these three CO lines are shown in Figure \ref{fig_coall}. The line fluxes and luminosities of all emission lines are listed in Table \ref{tbl_coline}. 

\begin{deluxetable*}{lccccccccccc}[tbh]
\tabletypesize{\scriptsize}
\tablecaption{Line Properties of J0100+2802. \label{table_cont}}
\tablewidth{0pt}
\tablehead{\colhead{} & \colhead{[CII]} & \colhead{[CI]\tablenotemark{a}}& \colhead{CO(11-10)} & \colhead{CO(10-9)}& \colhead{CO(7-6)}& \colhead{CO(6-5)}& \colhead{CO(2-1)\tablenotemark{b}}}
\startdata
Line Flux (Jy km s$^{-1}$)& 3.64$\pm$0.22 & $<$0.18& 0.28$\pm$0.04 & 0.25$\pm$0.05 & 0.22$\pm$0.06 &0.26$\pm$0.05 & 0.038$\pm$0.013 \\
Line Luminosity (10$^8$ L$_\odot$)& 37.02$\pm$2.24 &$<$0.78 & 1.90$\pm$0.27 & 1.54$\pm$0.31 & 0.95$\pm$0.26 & 0.96$\pm$0.19 & 0.05$\pm$0.02\\
Line Luminosity (10$^{10}$ K km s$^{-1}$ pc$^2$)& 1.69$\pm$0.10 &$<$0.46 &0.29$\pm$0.04 & 0.32$\pm$0.06 & 0.57$\pm$0.15 & 0.91$\pm$0.17 & 1.20$\pm$0.41\\
\enddata
\tablenotetext{a}{3-$\sigma$ limits.}
\tablenotetext{b}{Line flux is measured from NSF's Karl G. Jansky Very Large Array (VLA) by \cite{Wangr16} .}
\label{tbl_coline}
\end{deluxetable*}

\subsection{Dust Continuum Emissions}\label{subsec_cont}
The line emission observations are also used to measure the underlying dust continuum emissions as shown from Figure \ref{fig_cii_blue_red} through Figure \ref{fig_coall}. Our observations at the [\Cii] setup provide the highest signal-to-noise continuum observations. In order to cover more frequency space, we generate three continuum maps from the [\Cii] observations. One is created by averaging the line-free channels in the upper side band. This map provides the measurements of the continuum flux density at the mean frequency of 258 GHz. The second map is constructed by averaging all the channels in the spectral windows in the lower side band which provides the measurements of the continuum flux density at the mean frequency of 244 GHz. The third map is constructed by averaging all line free channels, including both the lower side band and the upper side band. This map centers at 251 GHz and provides the most sensitive measurement of the dust continuum. The final continuum 1$\sigma$ rms sensitivities at 258 GHz, 251 GHz, and 244 GHz are 25$\mu$Jy/beam, 17$\mu$Jy/beam, 21$\mu$Jy/beam, respectively. The 251 GHz continuum map which has the highest signal-to-noise is shown in Figure \ref{fig_cii_blue_red}.

In order to measure the sizes of the dust emitting regions, we fit 2D Gaussian functions to the three continuum maps close to [\Cii] using the {\it imfit} routine. We derive a deconvolved size of ($0\farcs47\pm0\farcs06$)$\times$($0\farcs27\pm0\farcs04$) and the integrated continuum flux density of 1.04$\pm$0.10 mJy for the deepest 251 GHz continuum map. The sizes of 258 GHz and 244 GHz continuum emissions are measured to be ($0\farcs46\pm0\farcs08$)$\times$($0\farcs25\pm0\farcs06$), and ($0\farcs48\pm0\farcs08$)$\times$($0\farcs27\pm0\farcs05$), respectively. The integrated continuum flux densities at 258 GHz and 244 GHz are 1.08$\pm$0.15 mJy and 0.98$\pm$0.13 mJy, respectively. 

\begin{figure*}
\centering
\includegraphics[width=0.49\textwidth]{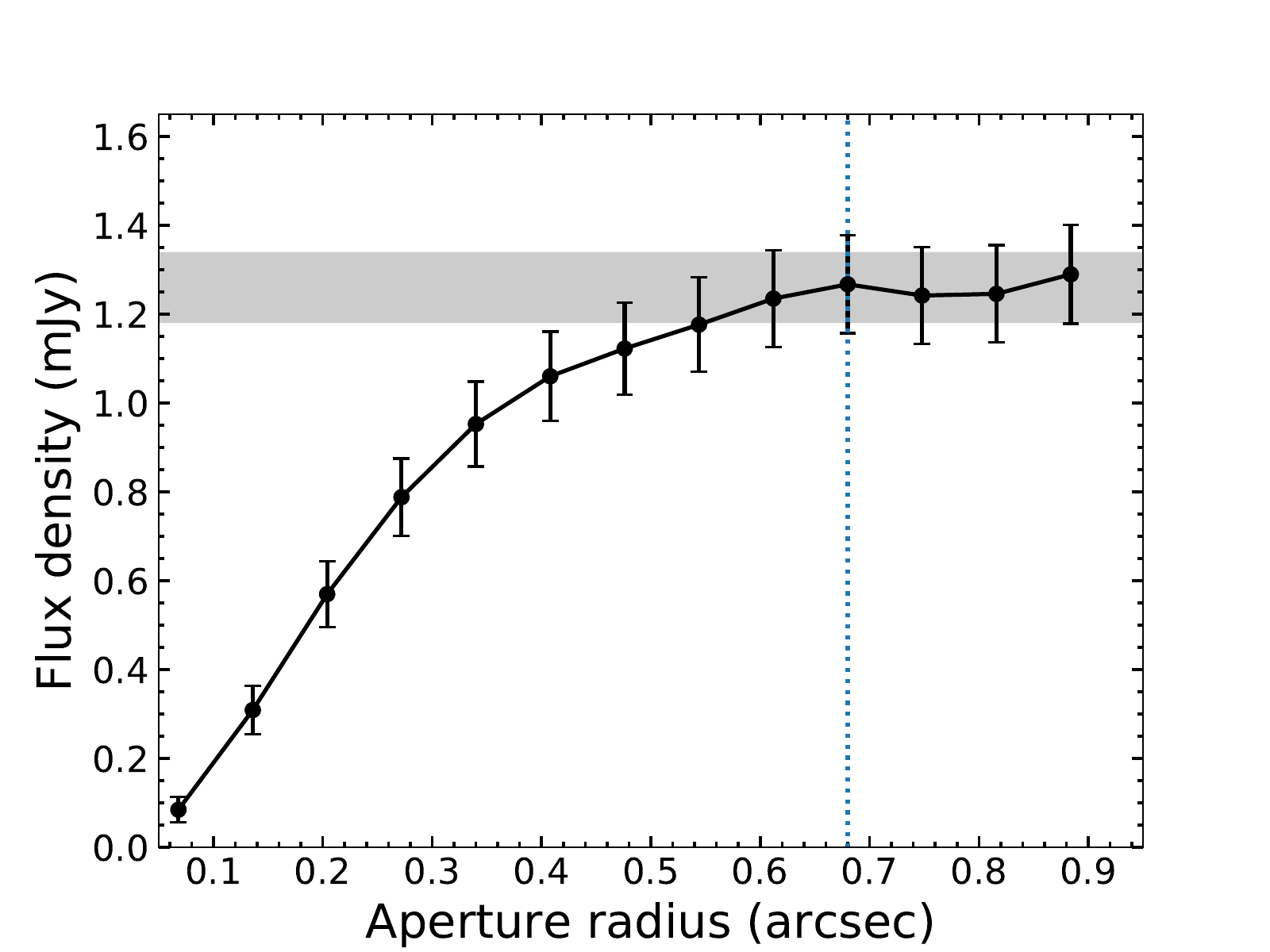}
\includegraphics[width=0.49\textwidth]{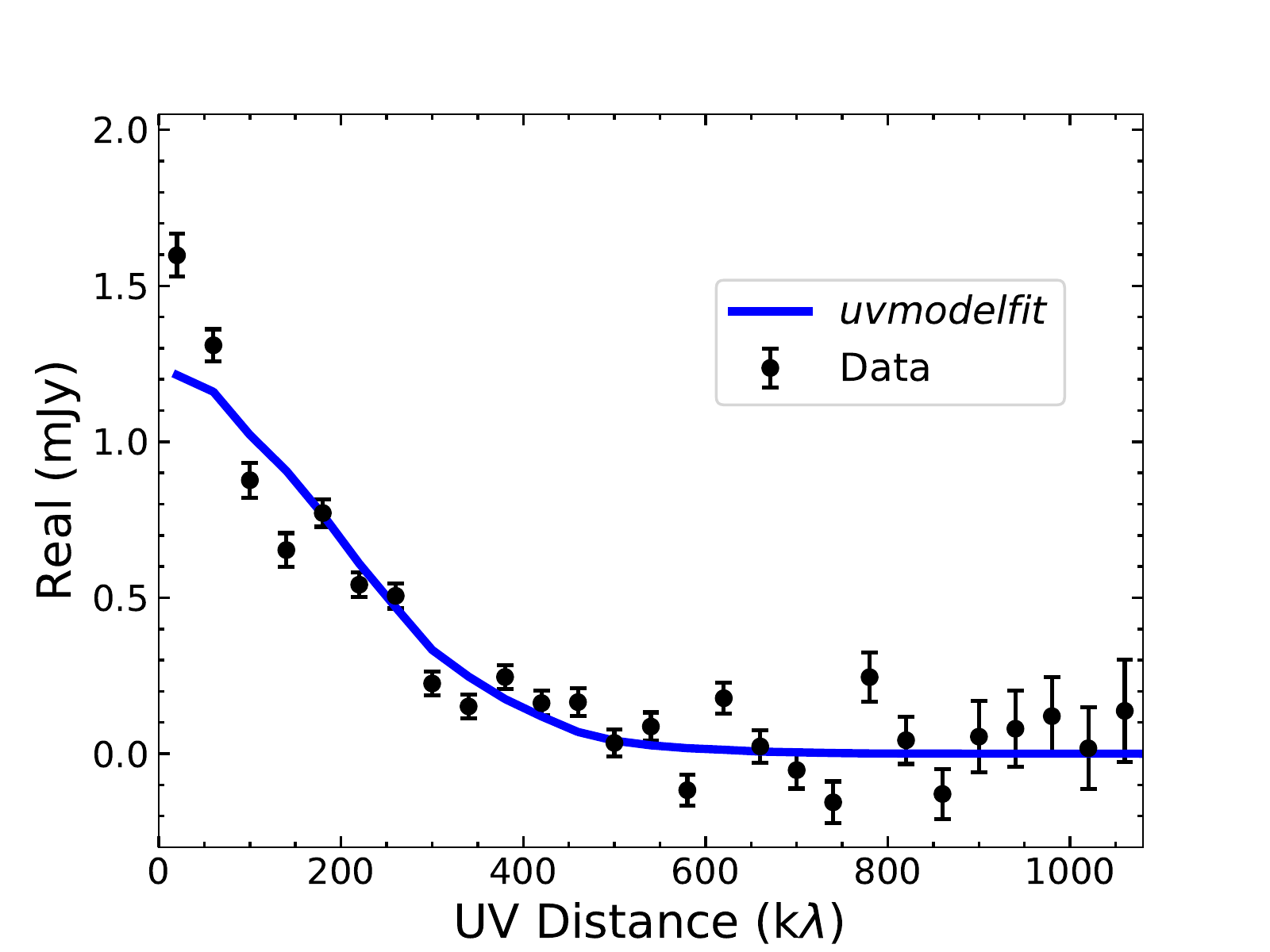}
\caption{
{\it Left}: The flux density of the 251 GHz continuum emission as a function of aperture radius. The flux density reaches a roughly constant value at radius $\gtrsim$ 0\farcs68. This is consistent with the peak flux density (grey shaded region) measured from the tapered continuum map.
{\it Right}: Visibility ({\it uv})-profile of the continuum emission from our band 6 observations. The visibility is radially averaged with a bin size of 40 $\rm k\lambda$ (black points). The blue solid line is the best-fit {\it uv} model which gives a size of FWHM=($0\farcs51\pm0\farcs02$)$\times$($0\farcs31\pm0\farcs02$) and a flux density of $1.23\pm0.05$ mJy at 251 GHz.
\label{fig_cont1}}
\end{figure*}

The continuum flux density measured around [\Cii] emission is only $\sim$80\% of that measured from unresolved (beam size is 2\farcs0 $\times$ 1\farcs7) PdBI observation \citep{Wangr16}, which suggests the continuum emission is well resolved. 
In order to measure the total continuum flux density (i.e. including faint extended emissions), we taper the continuum map with {\it uvtaper}=1\farcs5 in the {\it tclean} task. The peak flux densities measured from the tapered image are 1.47$\pm$0.10 mJy beam$^{-1}$, 1.26$\pm$0.08 mJy beam$^{-1}$ and 1.11$\pm$0.09 mJy beam$^{-1}$, for 258 GHz, 251 GHz, and 244 GHz, respectively.
Toward measuring the size of the missing extended continuum emission, we perform aperture photometry on the most sensitive continuum map at 251 GHz. We measure flux densities with a range of apertures and the growth curve is shown in the left panel of Figure \ref{fig_cont1}. The curve is flat at radius $\gtrsim$ 0\farcs68 with flux density consistent with the value observed at lower resolution \citep{Wangr16}. 

We also fit the {\it uv}-data of our band 6 continuum observations with the {\it uvmodelfit} task by assuming an elliptical Gaussian profile. The right panel of Figure \ref{fig_cont1} shows the {\it uv}-data and the best-fit {\it uv}-model profiles of the visibility which have been radially averaged in bins of 40 $\rm k\lambda$. The best-fit {\it uv}-model gives a total flux of $1.23\pm0.05$ mJy and a size of ($0\farcs51\pm0\farcs03$)$\times$($0\farcs31\pm0\farcs02$) or (2.83$\pm$0.11)$\times$(1.72$\pm$0.11) kpc$^2$ at 251 GHz. The flux is well consistent with what is measured from the tapered 251 GHz continuum image and the size is slightly bigger than that measured from the high resolution continuum image. We define the radius of the quasar host galaxy, $R_{\rm major}$, as the position where it is 3$\sigma$ (i.e. containing 99.7\% information) away from the line center along the major axis. Based on the best-fit {\it uv}-model, we measure $R_{\rm major}=0\farcs65\pm0\farcs04$ or $R_{\rm major}=3.6\pm0.2~{\rm kpc}$, which is consistent with both the growth curve measured from the continuum image and the size of the [\Cii] emission region. 
Since the 251 GHz has the highest sensitivity among our observations, we use the $R_{\rm major}$ derived here for constraining the dynamical mass of the quasar host galaxy in \S \ref{subsec_dynamics}.

We also generate continuum maps from each CO line observations by averaging line free regions in all four spectral windows. The final continuum 1$\sigma$ rms sensitivity at 99 GHz, 103 GHz, 152 GHz and 180 GHz are 4.3 $\mu$Jy/beam, 12 $\mu$Jy/beam, 17 $\mu$Jy/beam, and 19 $\mu$Jy/beam, respectively. The high resolution continuum map at 99 GHz is shown in Figure \ref{fig_co65} and other continuum maps with relative lower resolutions are shown in Figure \ref{fig_coall}.

We fit the continuum map around CO(6-5) with {\it imfit}. The image size (convolved with beam) is 0\farcs28$\times$0\farcs16, comparable to the beam size (0\farcs24$\times$0\farcs17), which suggests that the continuum emission at 99 GHz (at least the bright core) is unresolved. The peak flux density is measured to be 25$\pm$6 $\mu$Jy. We also tapered the continuum image at 99 GHz with {\it uvtaper}=1\farcs5, which gives a flux density at the peak position of 36$\pm$17 $\mu$Jy. The peak flux density from tapered image is about 1.4 times higher than that from the full resolution map suggests that we might missed some extended faint emissions in the high resolution map. Thus, we choose the peak flux density measured from the tapered image as the brightness at 99 GHz.

The continuum at 103 GHz is only marginally detected. Therefore,  instead of fitting the 2D continuum map, we adapt the peak flux density as the brightness, which is 43$\pm$12 $\mu$Jy. The continuum emission at 152 GHz is detected at $>$6$\sigma$ and we measure the flux density to be 178$\pm$19 $\mu$Jy for the bright core with {\it imfit}. 
We measure the continuum flux at 180 GHz to be 362$\pm$35 $\mu$Jy by using {\it imfit}. The continuum emissions at 103 GHz, 152 GHz and 180 GHz are not resolved (i.e. major axis of object is $<$1.2$\times$ that of beam) at the designed resolution. All continuum flux density measurements are listed in Table \ref{tbl_cont}.

J0100+2802 is the only quasar at $z>5.5$ that is detected by {\it GAIA} and thus provides us the  opportunity to investigate whether the position of the rest-frame UV AGN emission is offset from the (sub)-millimeter emissions. From Figure \ref{fig_cii_blue_red} to \ref{fig_coall}, we over-plot the optical position measured from {\it GAIA}. The {\it GAIA} position is fully consistent with the (sub)-millimeter emissions in both continuum and [\Cii] with no measurable offsets.

\begin{deluxetable*}{lccccccccccc}
\tabletypesize{\scriptsize}
\tablecaption{Continuum Properties of J0100+2802. \label{table_cont}}
\tablewidth{0pt}
\tablehead{
\colhead{$\nu_{obs}$ (GHz)} & \colhead{$\nu_{rest}$ (GHz)}& \colhead{Flux Density ($\mu$Jy)}
& \colhead{Beam Size}& \colhead{Deconvolved Size}& \colhead{Deconvolved Size (kpc$^2$)}}
\startdata
666\tablenotemark{a} & 4880 & $<$30000 &  7\farcs9$\times$7\farcs9 & -- & --\\
353\tablenotemark{a} & 2586 & 4100$\pm$1200 &  13\farcs0$\times$13\farcs0 & -- & --\\
258 & 1890 & 1470$\pm$100 & 0\farcs24$\times$0\farcs12 &0\farcs46$\times$0\farcs25 & 2.55$\times$1.39 \\
251 & 1839 & 1260$\pm$80 & 0\farcs25$\times$0\farcs13 &  0\farcs47$\times$0\farcs27 &  2.61$\times$1.50 \\
244 & 1788 &1110$\pm$90 & 0\farcs26$\times$0\farcs13 & 0\farcs48$\times$0\farcs27 & 2.66$\times$1.50 \\
180 & 1319 &362$\pm$35 & 0\farcs60$\times$0\farcs44 &  0\farcs40$\times$0\farcs28 & 2.22$\times$1.55 &\\
152 & 1114& 178$\pm$19 &0\farcs76$\times$0\farcs50 &0\farcs52$\times$0\farcs38 &  2.88$\times$2.11 & \\
103 & 755 & 43$\pm$12 &  0\farcs65$\times$0\farcs47 &  -- & -- & \\
99.4 & 728 &36$\pm$17 & 0\farcs24$\times$0\farcs17 &  -- & -- & \\
32\tablenotemark{a}  & 234& 14.8$\pm$4.3 & 0\farcs74$\times$0\farcs68 & -- & -- & \\
3\tablenotemark{a} &  22 &  104.5$\pm$3.1 & 0\farcs65$\times$0\farcs54 &  -- & -- & \\
1.5\tablenotemark{b} & 11 &91$\pm$17& 0\farcs012$\pm$0\farcs005 &0\farcs007$\pm$0\farcs003 &  0.04$\pm$0.02
\enddata
\tablenotetext{a}{Flux densities adopted from \cite{Wangr16}.}
\tablenotetext{b}{Adopted from \cite{Wangr17}.}
\label{tbl_cont}
\end{deluxetable*}

\section{Dust Emission Properties}\label{sec_dust}
\begin{figure}
\centering
\includegraphics[width=0.48\textwidth]{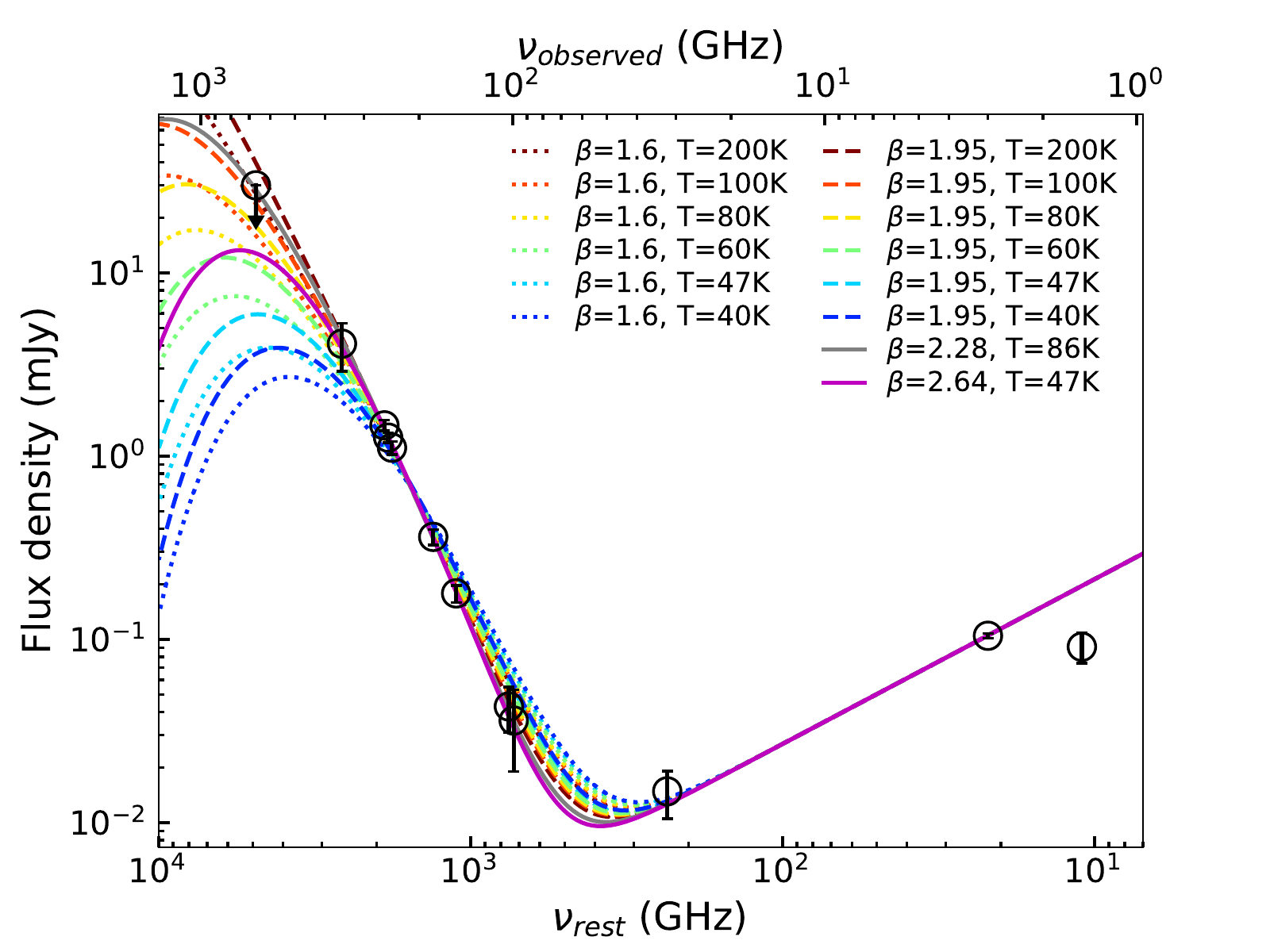} 
\caption{FIR dust continuum and radio emissions of the J0100+2802 host. Y-axis shows the measured flux densities listed in Table \ref{tbl_cont} and the x-axis shows the observed (top) and rest-frame (bottom) frequencies. The dashed lines are graybody model fits with the emissivity index fixed at $\beta$=1.95 \citep{Priddey01}, and the dotted lines are models with fixed $\beta$=1.6 \citep{Beelen06}.
 The magenta solid line denotes graybody model fitted with $\rm T_{dust}$ fixed at 47 K.  The grey solid line denotes graybody model fitted with both $\beta$ and $\rm T_{dust}$ are free parameters which yields $\beta$=2.28$\pm$0.26 and $\rm T_{dust}$=86$\pm$54 K.   
\label{fig_firsed}}
\end{figure}

\subsection{Dust Temperature and Emissivity Index }\label{subsec_dust}
Most high-redshift quasars lack full FIR spectral energy distribution (SED) measurements, and the $\rm L_{FIR}$ and $\rm M_{dust}$ are commonly determined with single or two photometric measurements by assuming an optically thin graybody model with a dust temperature of $\rm T_{dust}$=47K and an emissivity index of $\beta$=1.6 \citep[e.g.][]{Wangr16,Venemans18}. This temperature and emissivity index are measured from fitting a combined SED of a sample of high-redshift quasars with two or more rest-frame far-infrared photometric measurements by \cite{Beelen06}. However, both parameters have large scatters in different systems \citep[e.g.][]{Priddey01,Leipski13}. In this section, we combine the continuum measurements from \cite{Wangr16} and \cite{Wangr17} with our new ALMA observations to constrain the dust emission properties in J0100+2802. Before fitting the dust continuum, we subtract off the radio emissions determined by the 3 GHz and 32 GHz emission with the form of $f_\nu \propto \nu^{-0.9}$ \citep{Wangr16}. Note that the VLBA 1.5 GHz observation does not follow the steep spectra determined from the 3 GHz and 32 GHz observations which could be caused by the change of the spectra slope or the fact that the very high resolution VLBA observations could missed some extended flux as discussed in \cite{Wangr17}.

As suggested by previous works \citep[e.g.][]{daCunha13, Venemans16,Venemans17a}, it is important to take the effects of the cosmic microwave background (CMB) into account when studying high redshift objects. The CMB temperature at $z=6.3$ ($\rm T_{CMB}$ $\sim$ 20 K) is only slightly lower than dust temperatures in high-redshift quasar hosts, which reduces the flux densities that we can measure from sources at such high redshifts. We take the CMB effects into account following \cite{daCunha13}, who introduces a correction factor as a function of frequency and dust temperature at a certain redshift.

Note, however, most of the continuum points are on the Rayleigh--Jeans tail of the dust emission and we cannot tightly constrain the temperature and emissivity at the same time. Figure \ref{fig_firsed} shows the FIR dust continuum of the J0100+2802 host as well as different dust emission models with a set of $\beta$ and $\rm T_{dust}$ after taking into account the effects of the CMB as discussed above. We test a series of different graybody models. The different set graybody models as well as the radio continuum emission are shown in Figure \ref{fig_firsed}.
First, we set both dust emissivity index $\beta$ and dust temperature $\rm T_{dust}$ as free parameters which yields $\beta$=2.28$\pm$0.26 and $\rm T_{dust}$=86$\pm$54 K (grey solid line in Figure \ref{fig_firsed}). 

Since we do not have any continuum measurements at shorter wavelength, the $\rm T_{dust}$ has a large uncertainty. We then fix the temperature to be $\rm T_{dust}$=47 K and vary $\beta$ in the fitting which gives $\beta$=2.64$\pm$0.07 (magenta solid line in Figure \ref{fig_firsed}). The emissivity index is higher than the expected $\beta$=2 for pure silicate and/or graphite grains \citep[e.g.][]{Draine84}. However, such high emissivity index is also found in some local ultra-luminous infrared galaxies (ULIRGs) and can be explained by the presence of more than one single dust component \citep[e.g.][]{Clements10}. 

In order to better understand the allowed dust temperature and emissivity index in the host galaxy of J0100+2802, we further explore two different sets of graybody models in which $\beta$ is held fixed ($\beta$=1.6 and $\beta$=1.95), while the temperature is varied. These models are over-plotted in Figure \ref{fig_firsed}. By comparing those different models, we find that with slightly higher temperatures, models with $\beta$ fixed to 1.6 ($\rm T_{dust}\gtrsim$ 80 K) and 1.95 ($\rm T_{dust}\gtrsim$ 60 K) can also fit the data reasonably well. But the upper limit from the 450$\mu$m observation suggests that the dust temperature should not be higher than $\sim$100 K. The degeneracy between high $\beta$ and warm $\rm T_{dust}$ could be solved with ALMA band 9 and band 10 observations that reach the peak frequency of the SED. 

Since we only have rest-frame far-infrared continuum detections and cannot perform the multi-components fitting as investigated by \cite{Leipski13}. In order to further probe whether the temperature and emissivity index derived above affected by different fitting methods, we compare the observed continuum flux ratios of J0100+2802 with other luminous $z\gtrsim6$ quasars directly (i.e. model-free comparison). 
As most previous continuum observations of $z\gtrsim6$ quasars are mainly focused on frequencies close to [\Cii] and CO(6-5) lines, we derived the continuum flux density ratio $S_{\rm cont, [CII]}/S_{\rm cont, CO(6-5)}$= 35$\pm$17 for J0100+2802. Although $S_{\rm cont, [CII]}$ of many $z\gtrsim6$ quasars are detected \citep[e.g.][]{Venemans18}, only six quasars were solid detected (i.e., $>3\sigma$) for $S_{\rm cont, CO(6-5)}$ \citep{Wang10,Wang11,Venemans17b,Feruglio18} in the literatures to our knowledge. The $S_{\rm cont, [CII]}/S_{\rm cont, CO(6-5)}$ of these six quasars are in the range of 14.3--23.1, lower than that of J0100+2802. This is consistent with our fitting which suggests that the dust in the host galaxy of J0100+2802 has a high temperature and/or a large dust emissivity index than the average values of other $z\gtrsim6$ luminous quasars \citep[e.g.][]{Beelen06,Priddey01,Leipski13}. But we note that, there are some $z\gtrsim6$ quasars are not detected in the $S_{\rm cont, CO(6-5)}$ \citep[e.g.][]{Wang10,Bertoldi03} which makes them could have $S_{\rm cont, [CII]}/S_{\rm cont, CO(6-5)}\gtrsim30$, comparable to J0100+2802. Thus, we emphasis that future dust continuum observations at high frequency are needed to confirm whether the dust properties of J0100+2802 is different from other $z\gtrsim6$ quasars.

\subsection{FIR Luminosity and Dust Mass}\label{subsec_fir}
In this subsection, we describe the effects on FIR luminosity, dust mass and SFR measurements caused by the uncertainty of dust SED.
We have to assume a SED shape of the dust emission in order to compute the FIR luminosity. In previous works, the dust SED is often assumed to be a thin graybody with $\beta$=1.6 and $\rm T_{dust}$=47 K \citep{Beelen06}. With such assumption, we measured the FIR luminosity to be $L_{\rm FIR}$ = (3.5$\pm$0.7) $\times$ 10$^{12}$ $L_\odot$, the dust mass to be $M_{\rm dust}$=2.0$\times$10$^8$ M$_\odot$, and the SFR of 850 M$_\odot$ yr$^{-1}$ in \cite{Wangr16}. 

However, as shown above, the dust SED of J0100+2802 favors a warmer temperature and/or a higher emissivity index. 
Following \cite{Wangr16}, we derive the FIR luminosity to be $L_{\rm FIR}$ = 7.8$\times$ 10$^{12}$ $L_\odot$ and total infrared luminosity of $L_{\rm TIR}$ = 1.3$\times$ 10$^{13}$ $L_\odot$ by integrating the graybody with $\rm T_{dust}$=47 K and $\beta$=2.64 from 42.5$\mu$m to 122.5$\mu$m and from 3$\mu$m to 1100$\mu$m, respectively. Applying a scaling relation between $L_{\rm TIR}$ and the SFR found in the local universe: log SFR (M$_\odot$ yr$^{-1}$)= log $L_{\rm TIR}$ (erg s$^{-1}$)-43.41 \citep{Kennicutt12}, we find the SFR in J0100+2802 is 1900 M$_\odot$ yr$^{-1}$.
Previous studies of high redshift starburst galaxies show that the dust emission could be optically thick at $\lambda_{\rm rest}\lesssim200 \mu m$ \citep[e.g.][]{Riechers13}, thus using continuum measurements at wavelength significantly longer than $200 \mu m$ are preferred to estimate the dust mass. Since the emission at $\lambda_{\rm rest}>500\mu m$ of J0100+2802 could be affected by the non-thermal radio emission (see Figure \ref{fig_firsed}), we thus use the CO (6-5) underlying continuum to estimate the dust mass.
We follow the procedure described in \cite{Venemans18} by assuming the dust opacity coefficient $\kappa_{\nu}(\beta)=0.77~(\nu/352~{\rm GHz})^\beta$ cm$^2$ g$^{-1}$. The dust mass estimated based on the above dust temperature and emissivity index assumption is $M_{\rm dust}$=3.4$\times$10$^7$ M$_\odot$.

Assuming a graybody with $\rm T_{dust}$=60 K and $\beta$=1.95, we obtain $L_{\rm FIR}$ = 6.8$\times$ 10$^{12}$ $L_\odot$, $L_{\rm TIR}$ = 1.4$\times$ 10$^{13}$ $L_\odot$, SFR=2060 M$_\odot$ yr$^{-1}$ and $M_{\rm dust}$=5.8$\times$10$^7$ M$_\odot$. If a higher dust temperature of 80 K and $\beta$=1.6 adopt, these values would be $L_{\rm FIR}$ = 8.0$\times$ 10$^{12}$ $L_\odot$, $L_{\rm TIR}$ = 2.5$\times$ 10$^{13}$ $L_\odot$, SFR=3760 M$_\odot$ yr$^{-1}$, and $M_{\rm dust}$=5.8$\times$10$^7$ M$_\odot$. The FIR luminosities and SFRs derived above are about 2--4 times higher and the dust masses are about 2--4 times lower than that derived by simply assuming $\beta$=1.6 and $\rm T_{dust}$=47 K. 

As suggested by \cite{Clements10}, a steep dust SED can be explained with more than one dust components. 
In the discussion above, we assume that dust in J0100+2802 is heated mainly by star formation, which allows the estimation of SFR based on FIR luminosity. 
However, part of the dust emission in J0100+2802, the most intrinsically luminous object found at $z>6$,  might also be heated directly by the central power AGN. Our analysis on the gas excitation in the following section also supports this scenario, which will introduce additional uncertainties on estimating the SFR and dust mass in the host galaxy of J0100+2802. Therefore, we caution that the FIR luminosity, SFR and dust mass measured here are still highly uncertain and additional photometry at shorter wavelength is needed to better constrain these parameters for the host galaxy of J0100+2802.

\section{Gas Excitation}\label{sec_excitation}
\subsection{CO Spectral Line Energy Distribution}\label{subsec_cosled}

\begin{figure}
\centering
\includegraphics[width=0.48\textwidth]{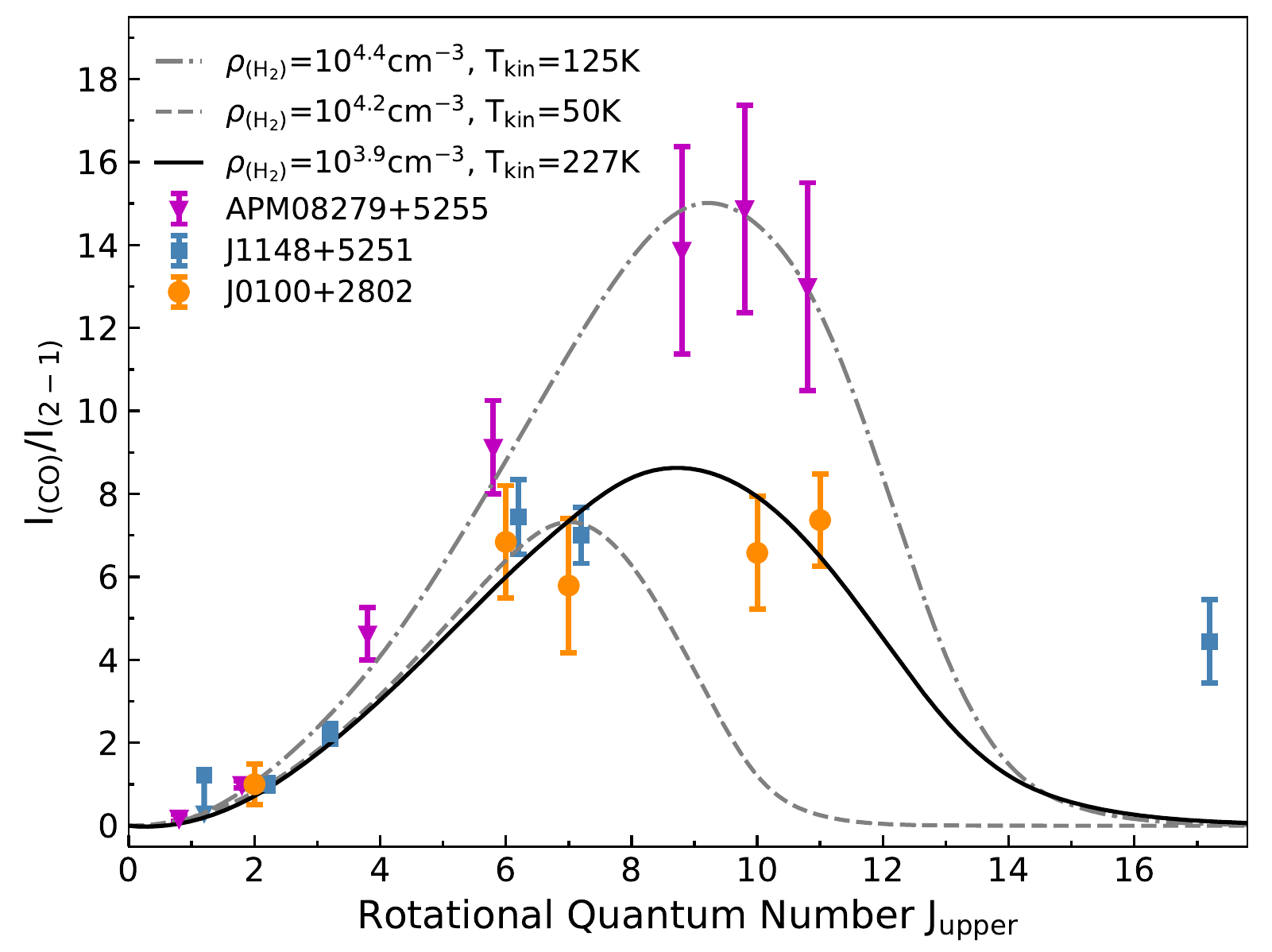} 
\caption{CO SLEDs of J0100+2802, J1148+5251 and APM 08279+5255. The orange circles denote measured CO SLED of J0100+2802, the steel blue squares and magenta triangles denote measured CO SLEDs of J1148+5251 and APM08279+5255, respectively. The gray dashed line denote single LVG model fitting for J1148+5251 from \cite{Riechers09} and the gray dash-dotted line denote single LVG model fitting for APM08279+5255 from \cite{Weiss07}. The black solid curve represents the best fit single LVG model for J0100+2802 with $n_{\rm (H_2)}$=10$^{3.9}$ cm$^{-3}$ and T$_{\rm kin}$=227 K.
\label{fig_cosled}}
\end{figure}

Low-J CO emissions provide crucial information on the cold molecular gas; on the other hand, high-J CO lines  have critical densities $\rm>10^5~ cm^{-3}$ and trace the warmer molecular gas in the center of galaxies. Thus the spectral line energy distribution (SLED) of the CO emission lines can reveal the physical conditions, especially the gas temperature and density, of molecular gas \citep[see][for a review]{Carilli13}. One of the most detailed analysis example is the highly excited gravitationally lensed quasar (APM 08279+5255) host galaxy at $z=3.9$, investigated by \cite{Weiss07}. In that study, they found that APM 08279+5255 can be modeled with a single-component large velocity gradient (LVG) model with H$_{\rm 2}$ density of $n_{\rm (H_2)}$=10$^{4.4}$ cm$^{-3}$ and a gas temperature of T$_{\rm kin}$=125 K. The fitting can be improved if they introduce both a cool gas component ($\sim$65 K) powered by star formation and a warm ($\sim$ 220 K) gas component directly heated by the central AGN. At higher redshifts (i.e. $z>6$), the only CO SLED of quasar host galaxy that has been studied in details is that of the quasar J1148+5251 at $z=6.42$ \citep{Riechers09,Gallerani14}. \cite{Riechers09} found the CO SLED of J1148+5251 host galaxy can be best described by a single LVG model with T$_{\rm kin}$=50 K, and $n_{\rm (H_2)}$=10$^{4.2}$ cm$^{-3}$. Recently, \cite{Gallerani14} claim the detection of a very high-J CO (17-16) emission and suggest that the CO excitation can be explained by introducing a composite of Photodissociation and X-ray-dominated region (PDRs and XDRs) models. 

\begin{figure}
\centering
\includegraphics[width=0.48\textwidth]{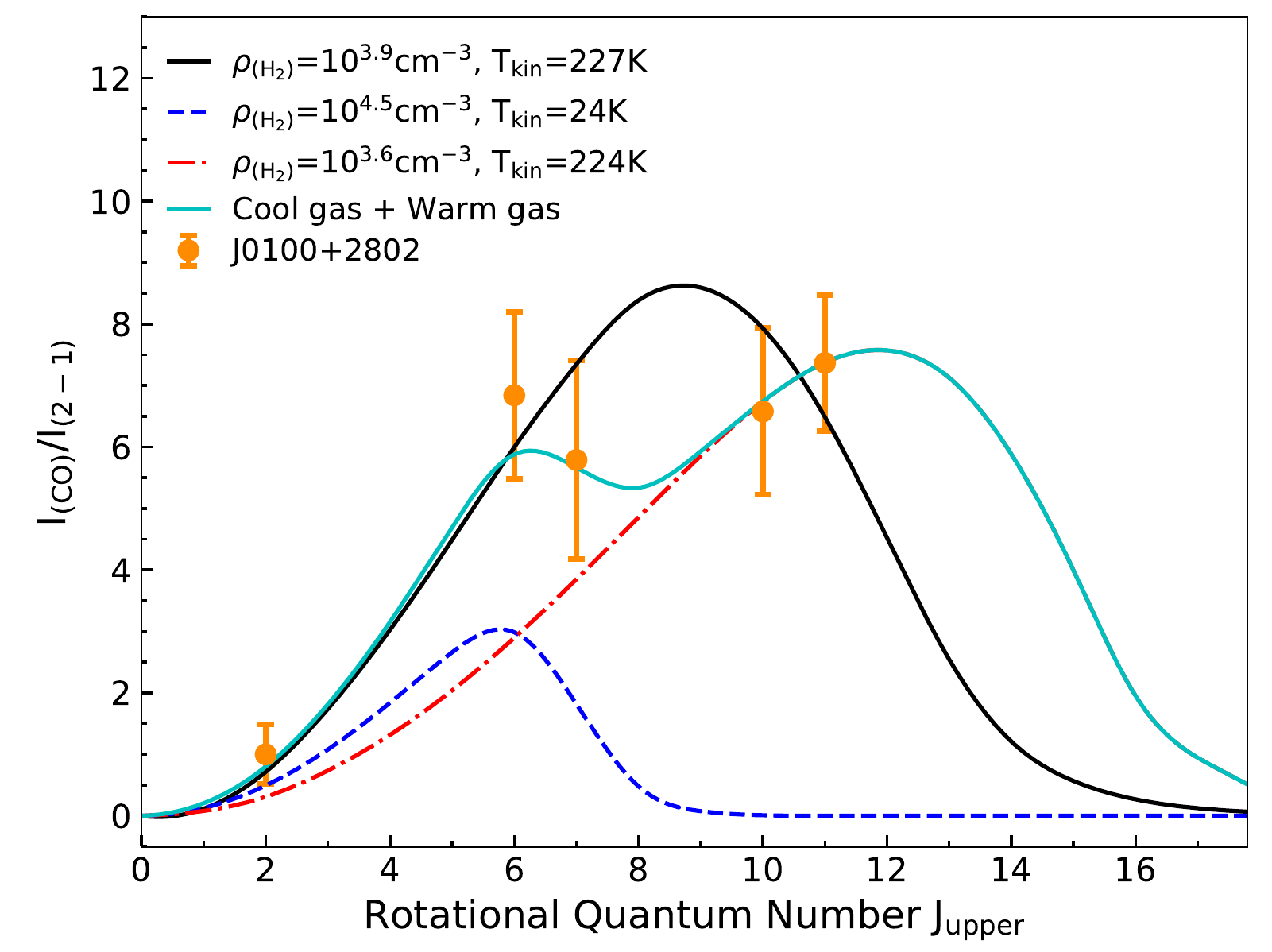} 
\caption{CO SLEDs of J0100+2802. The orange circles denote measured CO SLED of J0100+2802. The black solid curve represent best fit single LVG model for J0100+2802 with $n_{\rm (H_2)}$=10$^{3.9}$ cm$^{-3}$ and T$_{\rm kin}$=227 K. The blue dashed line and red dash-dotted lines represent cool ($n_{\rm (H_2)}$=10$^{4.5}$ cm$^{-3}$ and T$_{\rm kin}$=24 K) and hot ($n_{\rm (H_2)}$=10$^{3.6}$ cm$^{-3}$ and T$_{\rm kin}$=224 K) gas component, respectively. The cyan solid line is the sum of cool and hot gas components. 
\label{fig_cosled2}}
\end{figure}

Since we have detected five CO lines, it is possible to diagnose the gas excitation in J0100+2802. The CO SLEDs of J0100+2802 and two  examples mentioned above are shown in Figure \ref{fig_cosled}. We normalize the CO SLEDs of each system by their CO (2-1) emissions for comparison. We use a one-dimensional (1D) non-LTE radiative transfer code, RADEX, developed by \cite{VanderTak07} to study the CO excitation in the host galaxy of J0100+2802. The inputs of RADEX are the gas kinetic temperature (T$_{\rm kin}$), the volume density of the molecular hydrogen ($n_{\rm (H_2)}$), and the column density of the CO molecule ($\rm N_{\rm CO}$).  
We set the background temperature to be the CMB temperature at $z=6.327$. We searched the minimum ${\chi}^2$ within the parameter spaces of $n_{\rm (H_2)}$=10$^2$--10$^7$ cm$^{-3}$, T$_{\rm kin}$=T$_{\rm CMB}$-10$^3$ K, $\rm N_{\rm CO}$/$dv$ = 10$^{15.5}$--10$^{19.5}$ cm$^{-2}$ (km s$^{-1}$)$^{-1}$ \citep{Yang17}. The minimum reduced ${\chi}^2$ (${\chi_{\rm red}}^2$=2) yields T$_{\rm kin}$=227 K, $n_{\rm (H_2)}$=10$^{3.9}$ cm$^{-3}$ for a single LVG model. The best fit model is shown as the black solid line in Figure \ref{fig_cosled}; the best-fit single LVG model does not  reproduce the shape of CO SLED well, suggesting the CO excitation in J0100+2802 is more complicated (i.e. needs two or more gas components). 

Similar to \cite{Weiss07} and \cite{Yang17}, we split the CO SLED into two components in the RADEX modeling. We adopt the same parameter space to that we used for single LVG model fitting. Instead of using ${\chi}^2$ fitting, we performed the Markov chain Monte Carlo \citep[MCMC;][]{mcmc} calculations to fit our observed fluxes with the fluxes generated from RADEX models following the procedure explored by \cite{Yang17}. The median value and $\pm$1$\sigma$ range of the values from the probability distribution are found to be  $n_{\rm (H_2)}$=10$^{4.5_{-1.1}^{+1.1}}$ cm$^{-3}$ and T$_{\rm kin}$=24$_{-3}^{+8}$ K for the cool component and $n_{\rm (H_2)}$=10$^{3.60_{-0.8}^{+1.3}}$ cm$^{-3}$ and T$_{\rm kin}$=224$_{-100}^{+165}$ K for the warm component. 

We measure that $L_{\rm CO(10-9)}/L_{\rm CO(6-5)}=1.60\pm0.28$ and $L_{\rm CO(11-10)}/L_{\rm CO(6-5)}=1.98\pm0.24$ in J0100+2802, which are much higher than that of most starburst galaxies (see Figure 6 in \cite{Carniani19}). 
This suggests that the central AGN in J0100+2802 could directly heat the molecular gas in the quasar host galaxy, especially considering that J0100+2802 is the most luminous AGN that hosting the most massive SMBH known at $z>6$. 
However, we note that there are some dusty starburst galaxies in the early Universe, which are not known to host luminous AGNs, also show comparable bright high-$J$ CO emissions that could be produced by less warm but higher density gas components \citep[e.g.][]{Riechers13,Yang17}. Thus, future ALMA observations of higher-J CO lines (i.e. $J_{\rm upp}>13$) are needed to distinguish whether the bright high-$J$ CO emissions are excited by powerful AGN or from less warm but higher density gas.

\subsection{Molecular Gas Mass}\label{subsec_gasmass}
For high-redshift FIR luminous objects, a luminosity-to-mass conversion factor of $\alpha_{\rm CO}$=0.8 $M_\odot$ (K km s$^{-1}$ pc$^2$)$^{-1}$, derived from moderate-density warm inter-cloud medium, is usually adopted \citep{Downes98} when calculating $\rm H_2$ mass from CO observations. For J0100+2802, the $\rm H_2$ density derived for cool and hot gas components are $n_{\rm (H_2)}\sim$10$^{4.5}$ cm$^{-3}$, and $n_{\rm (H_2)}\sim$10$^{3.6}$ cm$^{-3}$, respectively. This is comparable with that of ULIRGS which is $n_{\rm (H_2)}\sim$10$^{4.0}$ cm$^{-3}$.  
But the temperature of the warm gas components is about five times higher than that in ULIRGS. The conversion factor $\alpha_{\rm CO}$ scales as $n_{\rm (H_2)}^{0.5}/T_{kin}$ \citep{Weiss07}. 
This suggests the conversion factor for the warm gas component could be $\alpha_{\rm CO}$$\sim$0.16 $M_\odot$ (K km s$^{-1}$ pc$^2$)$^{-1}$. The CO(1-0) line fluxes estimated for cool gas and warm gas based on the best fit two-components LVG model are 6.0$\times$10$^{9}$ K km s$^{-1}$ pc$^2$ and 3.7$\times$10$^{9}$ K km s$^{-1}$ pc$^2$, respectively. Thus, we estimate the molecular gas mass to be $M_{\rm cool}$=($4.8\pm1.6$)$\times$10$^9$ $M_\odot$,  and $M_{\rm warm}$=(5.9$\pm$0.8)$\times$10$^8$ $M_\odot$ by assuming $\alpha_{\rm CO}$=0.8 $M_\odot$ (K km s$^{-1}$ pc$^2$)$^{-1}$ for the cool gas and $\alpha_{\rm CO}$=0.16 $M_\odot$ (K km s$^{-1}$ pc$^2$)$^{-1}$ for the warm gas, respectively. The total molecular gas mass is $M_{\rm H_2}$=($5.4\pm1.6$)$\times$10$^9$ $M_\odot$. The gas mass measured here is about two times lower than that measured by \cite{Wangr16}. This is because that \cite{Wangr16} adopt $\alpha_{\rm CO}$=0.8 $M_\odot$ (K km s$^{-1}$ pc$^2$)$^{-1}$ for the total CO(1-0) flux and assumed that  $L_{\rm CO(1-0)}^{'}$ = $L_{\rm CO(2-1)}^{'}$.
Since the CO SLED model of J0100+2802 is still highly uncertain, and the measurable line luminosities from the cool component could be affected by the CMB which has comparable temperature with that of the cool component, the estimated gas mass still has large uncertainties.

On the other hand, one can also estimate the gas mass from the dust mass \citep[e.g.][]{Draine07}. From our dust SED models, we derive a dust mass of $M_{\rm dust}$=3.4$\times$10$^7$ M$_\odot$ for $\rm T_{dust}$=47 K and $\beta$=2.64, $M_{\rm dust}$=5.8$\times$10$^7$ M$_\odot$ for $\rm T_{dust}$=60 K and $\beta$=1.95 and $M_{\rm dust}$=5.8$\times$10$^7$ M$_\odot$ for  $\rm T_{dust}$=80 K and $\beta$=1.6. Assuming a gas-to-dust mass ratio of 80 \citep[e.g.][]{Riechers13}, we obtain a gas mass of $M_{\rm H_2}$=2.7$\times$10$^9$--4.6$\times$10$^9$ $M_\odot$. This gas mass is similar to the one we estimated from the LVG model. 
Adopting the gas mass measured from the LVG model and SFR from dust SED fitting, we can derive the molecular gas depletion timescale: 
$t_{\rm dep}({\rm H_2}) = \frac{M_{\rm H_2}}{\rm SFR}$. By considering all the three different dust SEDs we explored in section \ref{subsec_dust}, we estimate the molecular gas depletion timescale in the host galaxy of J0100+2802 is only of order 10$^6$ years. However, given the high luminosity of the quasar it is possible that some fraction of the dust is heated by the central AGN directly as discussed in section \ref{subsec_dust}, in which case the gas depletion timescale quoted above becomes a lower limit.

\begin{figure}
\centering
\vspace{5pt}
\includegraphics[width=0.48\textwidth]{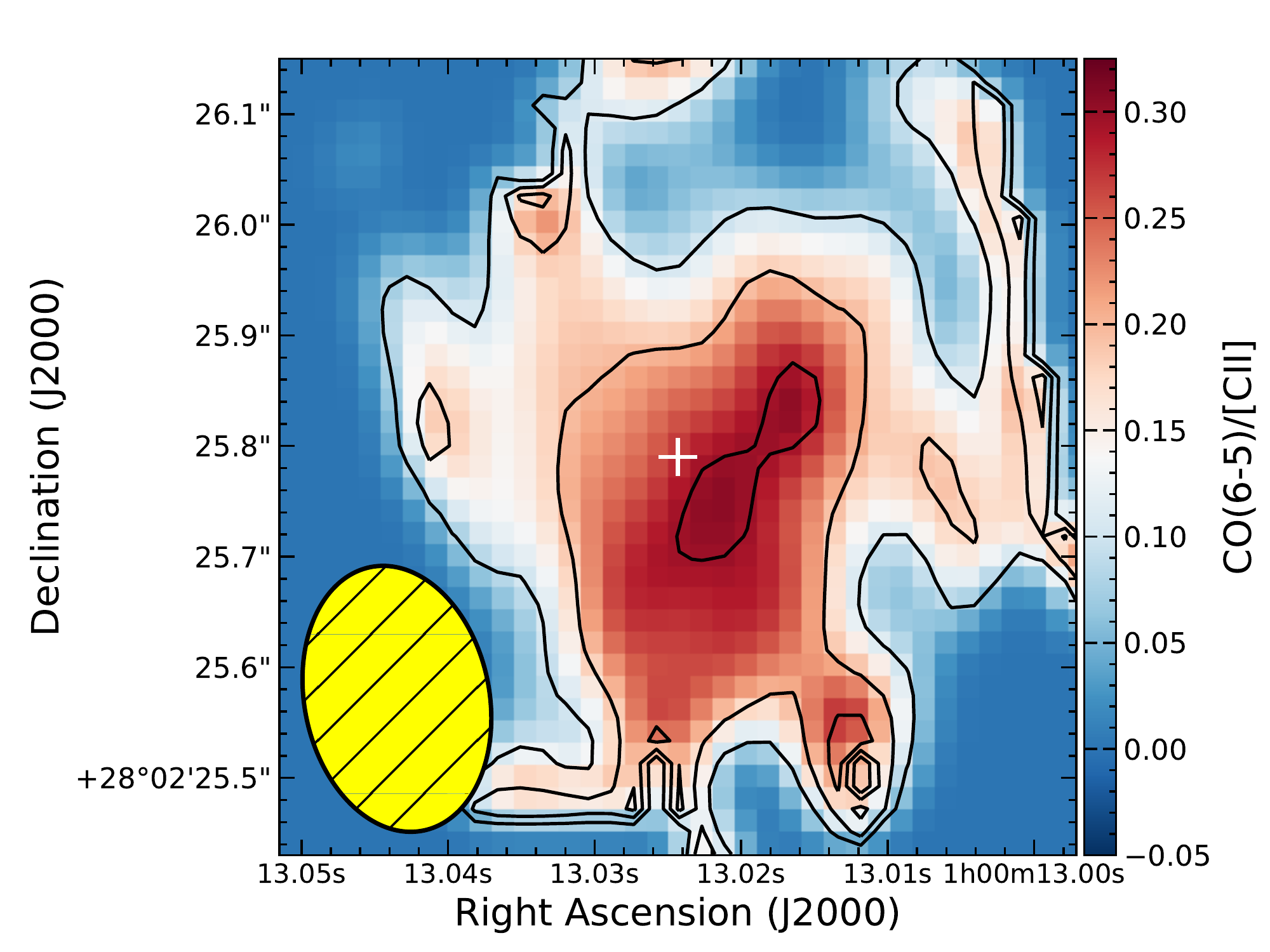} 
\caption{The CO(6-5) and [\Cii] line emission intensity map ratio within the central region (r=0\farcs4 or 2.2 kpc). The resolution for the [\Cii] intensity map is downgrade to be the same with that of CO(6-5). Both lines are integrated from --300 km s$^{-1}$ to +300 km s$^{-1}$. The beam is shown in the lower left corner as a cyan ellipse. The contours indicate the 0.1, 0.2, and 0.3 isophotes. 
The intensity ratio map suggests that dense molecular gas is more centrally concentrated with respect to the [\Cii] emission. The large ratio region at the edge of this map might be not reliable considering that both the [\Cii] and CO(6-5) intensities at these region are lower than 3$\sigma$.
\label{fig_co_cii_ratio}}
\end{figure}

\subsection{Line Ratios}\label{subsec_ratio}

\begin{figure*}
\centering
\includegraphics[width=0.95\textwidth]{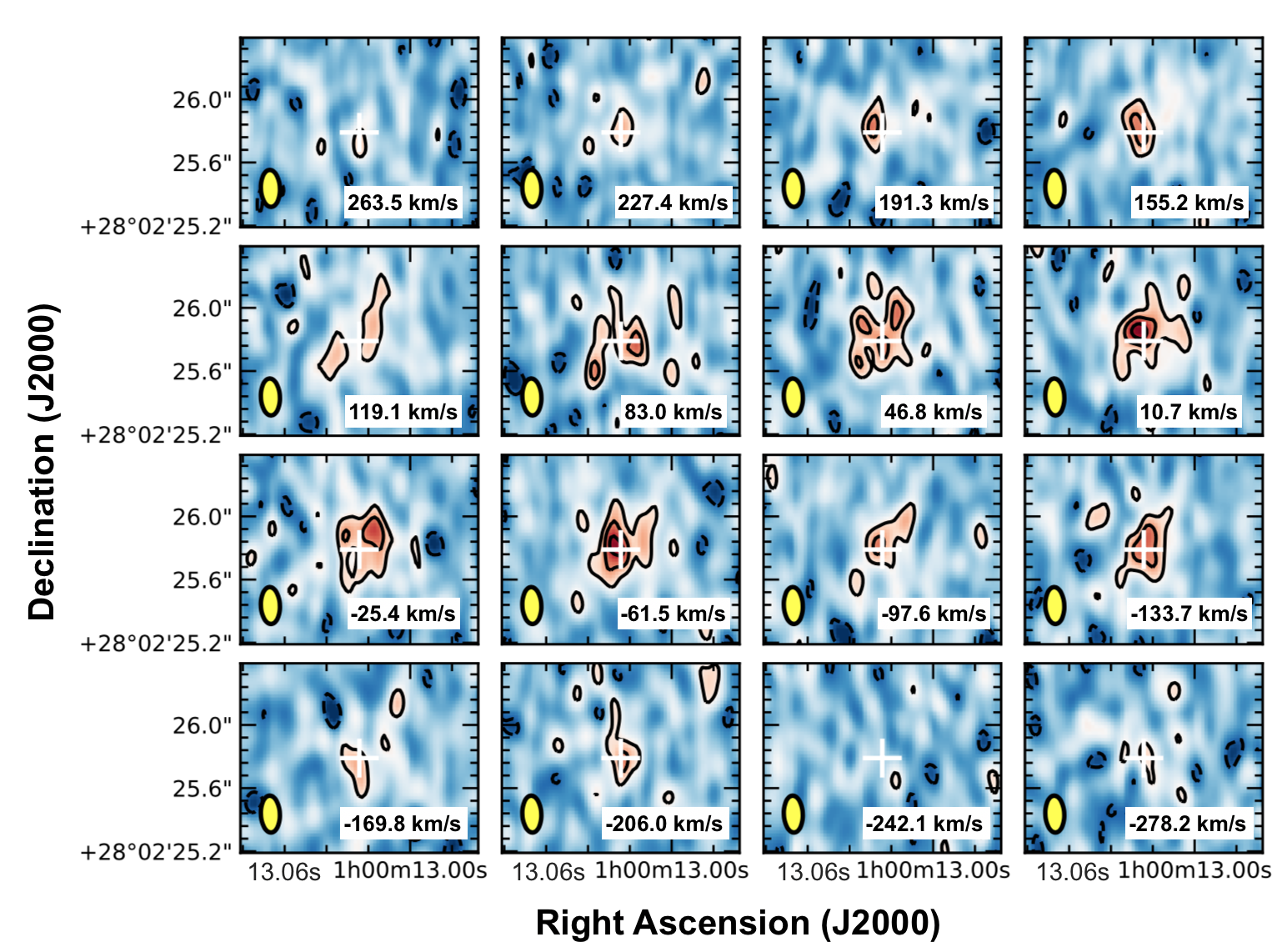}
\caption{[\Cii] emission channel map. The velocity resolution is $\sim$36 km s$^{-1}$. The [\Cii] emission is clearly detected in more than ten channels. Solid contour levels are +3, +5, and +7$\sigma$, where $\sigma$=0.21 mJy/beam. The --2$\sigma$ contours are presented by dashed lines. 
\label{fig_cii_velocity}}
\end{figure*}

In order to give a first order constraining on the spatially resolved excitation, we downgrade the resolution of the [\Cii] intensity map to be exactly the same with that of CO(6-5) emission. The intensity map ratio is shown in Figure \ref{fig_co_cii_ratio}. The intensity maps of both CO(6-5) and [\Cii] are integrated from --300 km s$^{-1}$ to +300 km s$^{-1}$. 
The CO(6-5)/[\Cii] ratio decreases from $\sim0.3$ to $\sim0.1$ from the galaxy center to regions out to $\sim1~{\rm kpc}$ indicating that: 1) the CO(6-5) is more centrally concentrated than [\Cii], 2) the overall gas density to be $n\gtrsim10^{4.5}$ cm$^{-3}$, unless the radiation field intensity is less than 1000 G$_0$ for a classic PDR model \citep{Kaufman99, Kaufman06, Pound08}, 3) the gas density and/or the radiation field are higher in the most central region than that at $\gtrsim1~{\rm kpc}$.
We also measure the [\Cii]/[\Ci] ratio as $\gtrsim$47, corresponding to a radiation field higher than $10^{3.5}$ G$\rm_0$ for $n\sim10^{5}$ cm$^{-3}$. All these line ratios suggest that both the gas density and the radiation field in the host galaxy of J0100+2802 are relatively high ($n\sim10^{5}$ cm$^{-3}$ and radiation field $>10^{3}~ {\rm G_0}$). 
The line luminosity ratio of [\Cii] and [\Ci] in J0100+2802 is significantly higher than the maximum line ratio predicted for XDR models of \cite{Meijerink07}. Similar results are also found in other $z>6$ quasars \citep[e.g.][]{Venemans17b}. This suggests that the atomic gas traced by [\Cii] and [\Ci] is dominated by the PDRs. 
Future high spatial resolution observations on multiple lines are necessary to further explore the spatially resolved excitation in the host galaxy of J0100+2802.

\begin{figure*}
\centering
\includegraphics[width=0.95\textwidth]{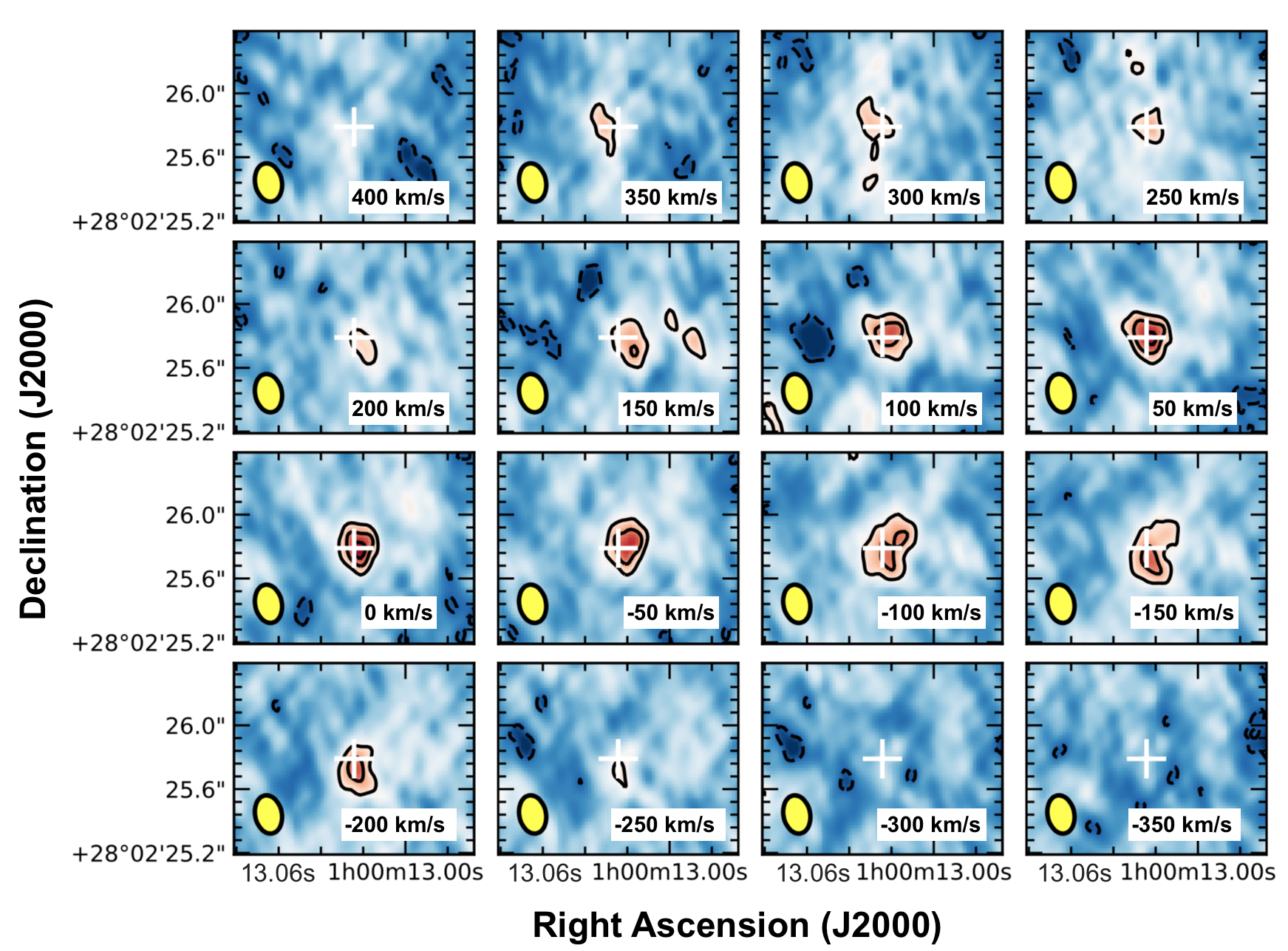}
\caption{CO(6-5) emission channel map. The velocity resolution is 50 km s$^{-1}$. The CO(6-5) is clearly detected in more than ten channels. Solid contour levels are +3, +5, and +7$\sigma$, where $\sigma$=0.058 mJy/beam. The --2$\sigma$ contours are presented by dashed lines. 
\label{fig_co_velocity}}
\end{figure*}

\section{Gas Kinematics and Dynamical Mass}\label{sec_kinematics}
\subsection{Gas Kinematics}\label{subsec_kinematics}
The coincide of the positions of blue and red sides of both CO(6-5) and [\Cii] emission lines suggests that both atomic and molecular gas do not have ordered motions on kpc scales. 
In order to further study the kinematics of the gas in the host galaxy of J0100+2802, we examine the velocity channel map of [\Cii] line in Figure \ref{fig_cii_velocity} and of CO(6-5) line in Figure \ref{fig_co_velocity}. The peak positions of both [\Cii] and CO(6-5) emissions in the individual velocity channels are fully consistent with the {\it GAIA} position, indicating that there is no clear rotation in the host galaxy of J0100+2802 on scales of  $\sim$1 kpc. Similar gas kinematics is also seen in other high-redshift quasar host galaxies \citep[e.g.][]{Walter09, Venemans17a} and brightest cluster galaxies \citep[BCGs; e.g.][]{McNamara14,Werner14}. 

If the [\Cii] and CO(6-5) structures are rotationally supported, its rotation axis must be very close to our line of sight (i.e. nearly face-on). The axial ratios derived from [\Cii] and CO(6-5) are $b/a=0.8\pm0.35$ and $b/a=0.74\pm0.67$, respectively. This could be consistent with a nearly face-on morphology although with large uncertainties. However, as shown in section \ref{sec_obs}, there could be additional extended emissions beyond our current detection limits, thus there could still be rotational components at larger scales. 

Figure \ref{fig_cii_velocity} suggests that there are some individual clumps in several velocity channels, but the size of these clumps are comparable to the beam size and the signal-to-noise ratio for those individual clumps detections is relative low (i.e. peaks at $\sim5\sigma$). Future deeper and higher resolution ALMA observations will allow us to reveal whether the host galaxy of J0100+2802 is clumpy or traced by mergers. 

\subsection{Dynamical Mass}\label{subsec_dynamics}
The dynamical mass of high redshift quasar host galaxy is usually estimated by assuming that the line is from a rotating disk. In \cite{Wangr16}, we measured the dynamical mass of J0100+2802 based on this method by assuming the diameter of the rotating gas disk to be $D=4.5\pm1.5$ kpc, the typical FWHM major axis sizes of [\Cii] emission from $z\sim$6 quasar hosts \citep[e.g.][]{Wangr13}. We had to make such assumption since we were not able to resolve the [\Cii] emission from our PdBI observations \citep{Wangr16}. 

In \S \ref{subsec_cont}, we measure the radius of the quasar host galaxy to be $R_{\rm major}$=3.6$\pm$0.2 kpc and the FWHM of [\Cii] emission to be 380$\pm$16 km s$^{-1}$. 
If the [\Cii] line is from a rotating gas disk and the circular velocity can be estimated as $v_{\rm cir}=0.75$ $\times$FWHM$_{\rm [CII]}$/sin($i$), where $i$ is the inclination angle ($i=0$ means face-on), the dynamical mass can then be estimated as $M_{\rm dyn}=1.16\times10^5~v_{\rm cir}^2~D$ = (7.0$\pm$1.3)$\times$10$^{10}$/sin$^2$($i$) $M_\odot$. 
The axis ratio of [\Cii] map is $b/a$=0.8$\pm$0.35,  corresponding to $i\sim37^\circ$. This yields a dynamical mass of $\sim1.9\times$10$^{11}M_\odot$,
an order of magnitude lower than that estimated based on the mass ratio of SMBHs and bulges in the local elliptical galaxies \citep[e.g., Equation 10 in ][]{Kormendy13}. 
However, since we do not see any velocity gradient in the [\Cii] emission and the inclination angle has large uncertainty, it is likely that  the gas is not supported by rotation, or that the disk is nearly face-on (i.e. $i\sim0^\circ$). If we assuming $i=5^\circ$, we would derive a dynamical mass of $\sim8.9\times$10$^{12}M_\odot$, comparable to that estimated from the local relation. 

\begin{figure}
\centering
\vspace{5pt}
\includegraphics[width=0.48\textwidth]{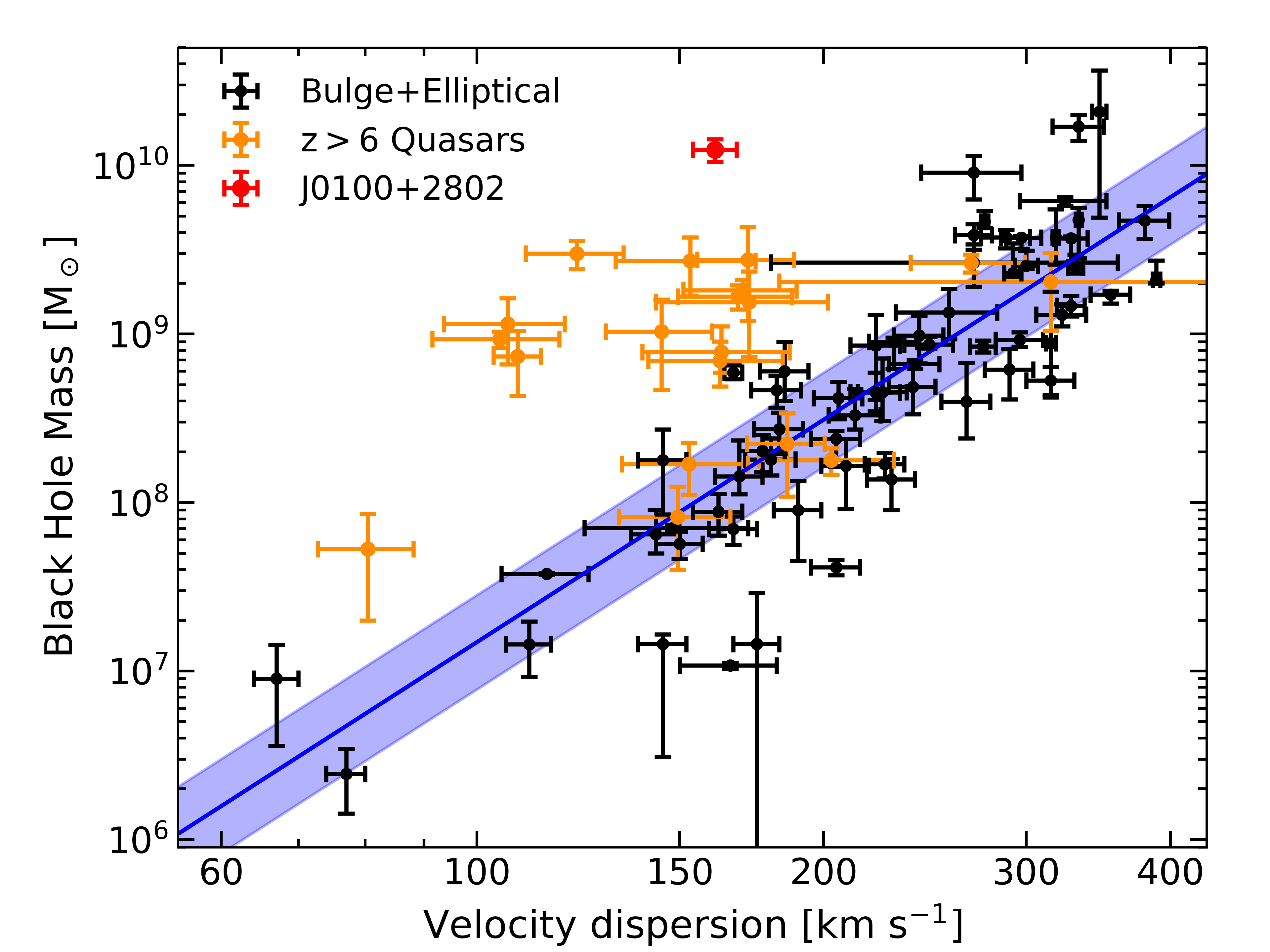} 
\caption{The $M_{\rm BH}$--$\sigma$ relation. The large red circle denotes J0100+2802. The orange circles are other $z\gtrsim6$ quasars with both \Mgii\ based single epoch virial BH mass measurements and [\Cii] line observations. The BH masses are compiled by \cite{Wang15} and [\Cii] line widths are from the compilation by \cite{Decarli18}. The small black circles are local elliptical galaxies or galaxies with classical bulges, collected by \cite{Kormendy13}. The blue solid line and shaded region denote the best fit $M_{\rm BH}$--$\sigma$ relation in local galaxies \citep{Kormendy13}. The systematic uncertainty of single epoch virial BH mass measurement which could be up to 0.5 dex \citep{Shen13} is not include in the plot.
\label{fig_msigma}}
\end{figure}

If the gas is dynamically hot and supported by random motion, we can estimate the dynamical mass use virial theorem following \cite{Venemans17a}. The velocity dispersion is measured to be 161$\pm$7 km s$^{-1}$ from the tapered spectrum shown in Figure \ref{fig_cii_blue_red}. The $M-\sigma$ relation of both $z>6$ quasars and local early type galaxies is shown in Figure \ref{fig_msigma}. Clearly, J0100+2802 as well as many other luminous high-redshift quasars are not following the $M_{\rm BH}$--$\sigma$ relation derived from local galaxies. The dynamical mass is measured to be $M_{\rm dyn}=3R\sigma^2/2G$=(3.25$\pm$0.46)$\times$10$^{10}$ M$_\odot$, which is comparable to that of other $z\sim$6 quasar host galaxies \citep[e.g.][]{Walter09, Wangr13, Willott15, Venemans16, Venemans17a, Venemans17c}. Intriguingly, the SMBH measured from single epoch virial method based on \Mgii\ line is 1.24$\times$10$^{10}$M$_\odot$ \citep{Wu15}, which is 38\% of the dynamical mass measured here. 
This could imply that J0100+2802 has the most large SMBH and dynamical mass ratio in all known $z\gtrsim6$ quasars with the next largest value (J1148+5251) to be $\sim$25\% \citep{Walter09,Willott15}. Even considering the large uncertainty on the SMBH mass measurement, which could be up to 0.5 dex \citep{Shen13}, the SMBH and dynamical mass ratio in J0100+2802 is still more than one order of magnitude higher than that in the local elliptical galaxies \citep[e.g.][]{Kormendy13}. 

\subsection{Mass Budgets}\label{subsec_mass}
The BH mass in J0100+2802 is measured to be $M_{\rm BH}$=(1.24$\pm$0.19)$\times$10$^{10}$ M$_\odot$ \citep{Wu15}. The molecular gas mass measured from the two components LVG model in \S \ref{subsec_gasmass} is $M_{\rm H_2}$=5.4$\times$10$^9$ $M_\odot$. We choose a fiducial dust SED of $\rm T_{dust}$=60 K and $\beta$=1.95, which gives a dust mass of $M_{\rm dust}$=5.9$\times$10$^7$ M$_\odot$. The atomic gas mass measured from [\Cii] emission is $M_{\rm atomic}$ $\sim$3$\times$10$^9$ M$_\odot$ \citep{Wangr16}. 
Put all these mass measurements together, we can give a first order constrain on the stellar mass in the host galaxy of J0100+2802. By assuming that dark matter does not significantly contribute to the mass budget \citep[e.g.][]{Genzel17}, we can give an upper limit of the stellar mass of $M_\ast<$1.2$\times$10$^{10}$ M$_\odot$ using the virial theorem measured dynamical mass.
But if we use the dynamical mass measured from a face on disk, the quasar host galaxy would contain a high stellar mass with order of 10$^{11}$ -- 10$^{12}$ $M_\odot$, which is at the high end of the stellar mass function derived for star-forming galaxies at similar redshifts \citep[e.g.][]{Bowler14}. The high stellar mass scenario is also supported by the presence of radio emission in J0100+2802 \citep{Wangr17}, since galaxies with more massive stellar mass tend to have more radio activities \citep[e.g.][]{Best05}. However, due to the large uncertainties in the derived dynamical mass, we can not put tight constraints on the stellar mass in the host galaxy of J0100+2802.

On another perspective, if we assume that the velocity dispersion measured from [\Cii] does represent the dynamics of the quasar host galaxy, we can use the velocity dispersion measured from [\Cii] emission line and the \Mgii\ based BH mass to estimate the radius of the sphere of influence of the central BH: $r_{\rm soi}=\frac{GM_{BH}}{\sigma_\ast^2} \sim 2$ kpc, which is more than two times larger than our beam size and comparable with the size of the [\Cii] emission region. This suggests that the kinematic of [\Cii] emission is strongly affected by the potential of central SMBH. This requires a re-examination of the mass budget discussed above. 
In this case, the mass estimated using the virial theorem can be interpreted as the upper limit of the BH mass if the [\Cii] emission dynamics is dominated by the SMBH potential.
On the other hand, if the host galaxy of J0100+2802 is a nearly face-on rotating disk as discussed above, the BH mass and dynamical mass could still follow the local relation \citep[e.g.][]{Kormendy13}. In this case, we can roughly estimate the radius of the sphere of influence of the central BH with the \Mgii\ based BH mass and the local $M_{\rm BH}-\sigma$ relation, which yields $r_{\rm soi} \sim 250$ pc. The sphere of influence is smaller than our beam size, but it is still much larger than that of other high-redshift quasars and can be resolved by future higher resolution ALMA observations and will allows us to dynamically measure the BH mass at the end of reionization.

Moreover, we note that the [\Cii] and CO(6-5) traced gas could be gravitationally unbounded, in which case the dynamical mass of the quasar host galaxy might be significantly different from the value quoted here. Although we do not see any obvious merger signatures in our current data, we can not rule out late stage merger that beyond the sensitivity of current observations. We also seem to miss some faint extended emissions and we can also not rule out minor mergers traced by faint galaxies at the outskirt of the quasar host galaxy. 

In summary,  we find that while the narrow [CII] and CO line width in the J0100+2802 host is suggestive of a modest dynamical mass compared to the prediction from local $M_{\rm BH}$ vs. $M_{\rm dyn}$ relation, observations with considerably higher S/N and spatial resolution are needed to carry out a full dynamical model by including the influence of BH gravity, pressure support from gas random motion, the possibility of a nearly face-on morphology, or the presence of merger activity.

\section{Summary} \label{sec_summary}
In this work, we present multi-band ALMA observations of dust continuum, [\Cii] emission and CO emission lines in the host galaxy of J0100+2802. Both dust continuum, [\Cii] emission and CO(6-5) are spatially resolved at sub-kpc scale. We also detected  high-J CO lines in CO(11-10), CO(10-9), and CO(7-6). The main findings of this work are listed as follows.
\begin{itemize}
\item[1.] We model the dust continuum SED over a wide frequency coverage and find that the the dust in J0100+2802 has either a high dust emissivity $\beta\gtrsim2$ or a high dust temperature $T_{\rm dust}\gtrsim60$ K or a combination of both factors. This distinguish J0100+2802 from other $z>6$ quasars most of which have lower dust temperatures. The FIR luminosity and SFR derived from several allowed dust SED models with warm temperature and/or high dust emissivity are about 2--4 times higher and the dust mass are about 2--4 times lower than that derived by simply assuming $\beta$=1.6 and $\rm T_{dust}$=47 K. 

\item[2.] We model the CO SLED of J0100+2802 with the LVG method. 
The CO SLED can be explained by a two gas components model, a cool component at $\sim24$ K with a high density of $n_{\rm (H_2)}$=10$^{4.5}$ cm$^{-3}$, and a warm component at $\sim224$ K with a slightly lower density of $n_{\rm (H_2)}$=10$^{3.6}$ cm$^{-3}$. 
The total molecular gas mass is measured to be $M_{\rm H_2}=5.4\times 10^9M_\odot$ based on the LVG model.
The high $L_{\rm CO(10-9)}/L_{\rm CO(6-5)}$ and $L_{\rm CO(11-10)}/L_{\rm CO(6-5)}$ ratios suggest that the central powerful AGN in J0100+2802 could directly heat the gas in the quasar host galaxy, but future ALMA observations of higher-J CO lines (i.e. $J_{\rm upp}>13$) are needed to distinguish whether the bright high-$J$ CO emissions are excited by powerful AGN or from less warm but higher density gas.

\item[3.] We also investigate the spatial distribution of [\Cii] and CO(6-5) line ratios in the host galaxy of J0100+2802, which indicates that the molecular gas traced by CO(6-5) emission lives in a more central concentrate dense region than the [\Cii] emission. 
With the bright [\Cii] emission, the non-detection of [\Ci] emission indicates that the atomic gas in the host galaxy of J0100+2802 is dominated by the PDRs. 

\item[4.] We examine the kinematics of J0100+2802 using spatially resolved [\Cii] and CO(6-5) observations, and find no ordered motion on kpc scales. 
The velocity dispersion measured from [\Cii] emission line is about three times smaller than that expected from the local $M_{\rm BH}-\sigma$ relation. 
The dynamical mass of J0100+2802 measured by using the virial theorem is only three times more massive than the central BH. However, our current observations can not rule out the galaxy to be a face on disk galaxy which still allow the dynamical mass of J0100+2802 following the local $M_{\rm BH}$ and $M_{\rm dyn}$ relation.

\end{itemize}

J0100+2802 is the most luminous object found at the end of reionization and the central BH mass is measured to be $1.24\times10^{10} M_\odot$ using the single-epoch virial method. The BH's gravitational radius of influence of J0100+2802 can be resolved with future considerably higher S/N and spatial resolution ALMA observations which will allow us to dynamically measure the BH mass. The notable highly excited molecular gas and warm dust temperature suggests that future deep ALMA observations on multiple emission lines and dust continuum are valuable for investigating the AGN feedback on the formation of early massive galaxy.  

\acknowledgments
F.W., X.-B.W. and R.W. acknowledge support from the National Key R\&D Program of China (2016YFA0400703) and the National Science Foundation of China (11533001 \& 11721303). X.F., J.Y., and M.Y. acknowledge support from the US NSF Grant AST-1515115 and NASA ADAP Grant NNX17AF28G. R.W. acknowledge support from the National Science Foundation of China grant No. 11473004. F.W. thanks ENIGMA group members at UCSB for valuable comments on this work. 
We thank the anonymous referee for carefully reading the manuscript and providing constructive comments and suggestions on improving the manuscript.
 
The National Radio Astronomy Observatory is a facility of the National Science Foundation operated under cooperative agreement by Associated Universities, Inc. This paper makes use of the following ALMA data: ADS/JAO.ALMA\#2015.1.00692.S and ADS/JAO.ALMA\#2017.1.00624.S. ALMA is a partnership of ESO (representing its member states), NSF (USA) and NINS (Japan), together with NRC (Canada), MOST and ASIAA (Taiwan), and KASI (Republic of Korea), in cooperation with the Republic of Chile. The Joint ALMA Observatory is operated by ESO, AUI/NRAO and NAOJ. This paper also used the data based on observations under projects VLA/14B-151 and VLA/15A-494 with the VLA, and project M15BI055 with JCMT/SCUBA-2.

\vspace{5mm}
\facilities{ALMA, VLA, JCMT (SCUBA-2)}

\end{document}